\pdfoutput=1
\documentclass[review]{elsarticle}

\usepackage{lineno}
\usepackage{graphicx}
\usepackage{subfig}
\usepackage{float}
\usepackage{url}
\usepackage{enumerate}
\usepackage{algorithm}
\usepackage{algpseudocode}
\usepackage{multirow}
\usepackage{bm}
\usepackage{pdflscape}
\usepackage{dsfont}
\usepackage[table]{xcolor}
\usepackage{amsmath,amsthm}
\usepackage[T1]{fontenc} 
\usepackage[utf8]{inputenc}
\usepackage{setspace}
\usepackage{picins}
\usepackage{hyperref}

\journal{Journal of Computer Communications}

\begin{document}

\begin{frontmatter}


\title{Power Interference Modeling for CSMA/CA based Networks using Directional Antenna}

\author[mainadd,secondadd]{Saravanan Kandasamy\corref{corauthor}}
\cortext[corauthor]{Corresponding author}
\ead{kandasamy@inesctec.pt}

\author[mainadd,secondadd]{Ricardo Morla}
\ead{ricardo.morla@inesctec.pt}

\author[mainadd,secondadd]{Manuel Ricardo}
\ead{mricardo@inesctec.pt}

\address[mainadd]{Centre for Telecommunications and Multimedia, INESC TEC, Porto, Portugal}
\address[secondadd]{Faculty of Engineering, University of Porto, Portugal}

\begin{abstract}
In IEEE 802.11 based wireless networks adding more access points does not always guarantee an increase of network capacity. In some cases, additional access points may contribute to degrade the aggregated network throughput as more interference is introduced.

This paper characterizes the power interference in CSMA/CA based networks consisting of nodes using directional antenna. The severity of the interference is quantized via an improved form of the $\textit{Attacking}$ $\textit{Case}$ metric as the original form of this metric was developed for nodes using omnidirectional antenna.

The proposed metric is attractive because it considers nodes using directional or omnidirectional antenna, and it enables the quantization of interference in wireless networks using multiple transmission power schemes. The improved $\textit{Attacking}$ $\textit{Case}$ metric is useful to study the aggregated throughput of IEEE 802.11 based networks; reducing $\textit{Attacking}$ $\textit{Case}$ probably results in an increase of aggregated throughput. This reduction can be implemented using strategies such as directional antenna, transmit power control, or both.
\end{abstract}


\end{frontmatter}


\section{Introduction}

IEEE 802.11 based wireless local area network (WLAN) technologies had a tremendous growth in recent years. Cheap and widely available equipments that can be deployed without a license are some of the factors contributing for the technology to gain popularity. A substantial number of access points (APs) are needed to provide coverage for areas such as a university or a city centre. Further, different entities may setup WLANs in the same geographical area uncoordinated. As a consequence, overlapping WLANs emerge. Lack of planning causes the network to saturate due to interference, and reach its capacity faster. Installing additional APs does not increase the capacity of network beyond a certain limit; moreover, if not done carefully the performance of the network could degrade further due to hidden and exposed nodes.

In wireless networks interference is a fundamental issue. Interference is the disturbance caused by a node's RF transmission into neighboring node(s). High transmission powers increase the number of nodes being interfered. The severity of interference can be quantized using the performance metric $\textit{Attacking}$ $\textit{Case}$ \cite{ivan}. This metric uses information such as nodes position, transmission power, signal to interference ratio and radio propagation model to characterize the instances where simultaneous transmissions are not allowed and, if allowed, the transmission would not be successful. A high $\textit{Attacking}$ $\textit{Case}$ value indicates a severe interference in the network. Therefore this metric is useful to understand and to optimize the performance of a wireless network.

The IEEE 802.11 standard caters for omnidirectional antenna (OA) \cite{mac} but there are many IEEE 802.11 based network deployed using directional antenna (DA) \cite{michael,david,eric,vibhav,roger}. The well known motivations for using DA \cite{sk,sk2} include: 1) a node is able to selectively send signals to desired directions. This allows the receiver node to avoid interference that comes from unwanted directions, thereby increasing the signal to interference and noise ratio (SINR); 2) more users could utilize a network simultaneously due to the spatial reuse factor which is higher than OA; 3) in a multihop network, a source is able to reach its destination node in a lesser number of hops due to the increase of transmission range because of the higher antenna gain. For these reasons, DA may be preferred to OA in some wireless network scenarios.

\begin{figure}
\centering
\includegraphics[clip,width=0.7\columnwidth]{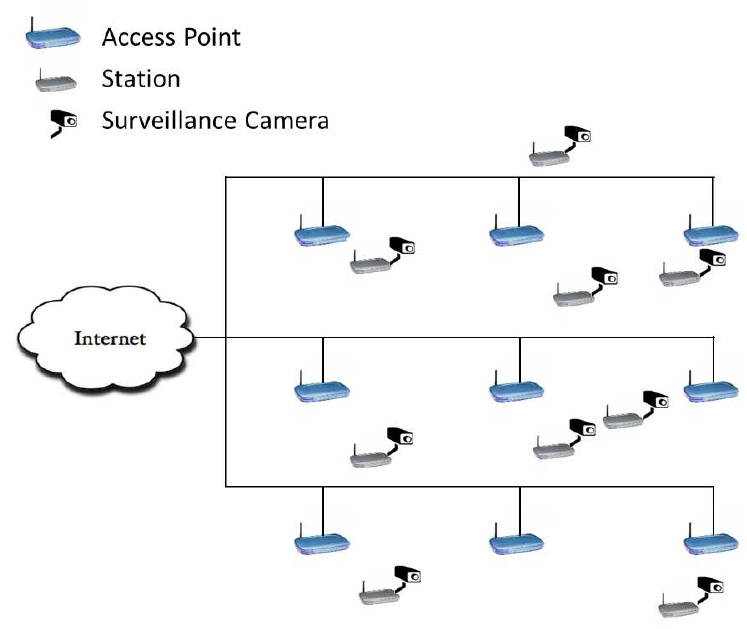}
\caption{The wireless videos surveillance network deployed as a basic scenario.} 		
\label{fig:topology}
\end{figure}

This paper aims to characterize the power interference for IEEE 802.11 based networks consisting of nodes using DA. To quantize the severity of interference in a wireless network, the $\textit{Attacking}$ $\textit{Case}$ metric defined in \cite{ivan} is adopted as reference and extended to cater for DA. The Link-Interference Graph, Transmitter-side Protocol Collision Prevention Graph, and Receiver-side Protocol Collision Prevention Graph are used to define the improved $\textit{Attacking}$ $\textit{Case}$ metric. Power constraints consisting of Physical Collision Constraints and Protocol Collision Prevention Constraints are utilized to model the graphs.

We have considered the wireless video surveillance network shown in Fig. \ref{fig:topology} as the basic scenario for our study. A video surveillance camera is attached to an IEEE 802.11 based station (STA) which is randomly placed in a network. The STA will connect to its closest AP placed at a fixed location and send its video traffic towards the AP. In our scenario the APs have access to the Internet via a wired connection. The network operates using the Basic Access Scheme of Distributed Coordinated Function (DCF) of the IEEE 802.11 MAC protocol known as Carrier Sense Multiple Access with Collision Avoidance (CSMA/CA). When a node (STA or AP) transmits, all other nodes within its power interference range are prohibited from transmitting in the same channel until the end of its current transmission. Individual DATA frames are acknowledged by an ACK frame and retransmission is scheduled by the sender if no ACK is received. Only when the medium is free the other nodes are allowed to transmit after waiting for a random time interval. As each STA is fitted with a video surveillance camera, it always has traffic to send and aggressively competes for accessing the medium. 

This paper provides one major contribution -
an improved $\textit{Attacking}$ $\textit{Case}$ metric that quantizes the severity of interference in IEEE 802.11 based networks consisting of nodes using DA. Our current metric differs from Liew's $\textit{Attacking}$ $\textit{Case}$ metric \cite{ivan} on the following aspects: a) the consideration of direction of transmission $\theta$ when the power constraints are built; b) the adoption of Protocol Collision Prevention Constraints using carrier sensing range and transmission range; c) association of a weight $w$ to the edge of the Link-Interference Graph, Transmitter-side Protocol Collision Prevention Graph, and Receiver-side Protocol Collision Prevention Graph. The improved $\textit{Attacking}$ $\textit{Case}$ is backward compatible with the former definition and can also be used in networks using OA. Our contribution can be particularly useful for network planners to understand the severity of interference in their network and make remedial actions to reduce it; an interference reduction effort is successful if ${Attacking\ Case}_{after} < {Attacking\ Case}_{before}$.

The rest of the paper is organized as follows. In Section 2 we present the related works and show the research space our work fills. In Section 3 we introduce the power constraints in IEEE 802.11 networks. In Section 4 we present the graph model used to obtain the improved $\textit{Attacking}$ $\textit{Case}$ metric. The power constraints are utilized to characterize the graph model. In Section 5 we describe the simulation carried out and the performance results obtained. Finally, in Section 6 we draw the conclusions and indicate topics for future work.

\section{Related Work}
\begin{figure}
\centering
\includegraphics[clip,width=0.7\columnwidth]{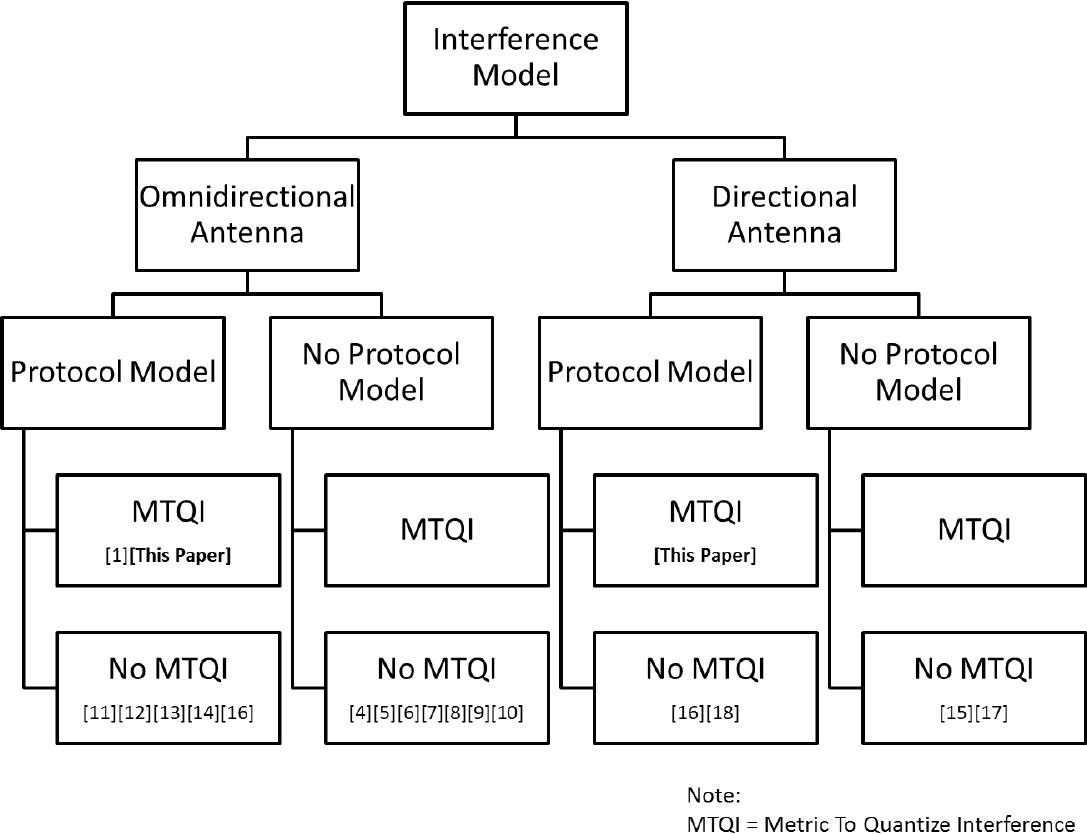}
\caption{Taxonomy for interference model} 
\label{fig:taxo}
\end{figure}

In this section we present relevant related works and review the literature from the perspective of interference modeling. Fig. \ref{fig:taxo} illustrates a possible taxonomy for interference models where the related works are categorized by antenna type, usage of protocol model, and proposal of a Metric To Quantize Interference (MTQI). This taxonomy will also be used to describe the research space our work fits in. 

The type of antenna a node uses influences the severity of interference in a wireless network. Renato and Fagner \cite{moraes} modeled interference for wireless ad hoc network; they found signal to interference plus noise ratio (SINR) approaches a constant value when the number of nodes increases around a receiving node if the path loss parameter is greater than two. Hence, communication is feasible for near neighbors though the number of interferers scales. Liu et al. \cite{liu} demonstrated the reduction of interference by tuning the carrier sense threshold; they concluded that the optimum carrier sensing range should be balanced with the spatial reuse and the impact of interference in order to optimize the aggregate throughput of nodes. The works by Renato and Fagner, and Liu et al. including several other recent works in \cite{zeng,varma,hoang,parvej} have modeled interference for nodes using OA and may not be suitable for nodes using DA. We modeled interference for nodes using DA and our proposed model does also address nodes using OA.

Gupta and Kumar proposed the Protocol Model \cite{pgupta}. Suppose $X_{i}$ refers to the physical position of node $i$. When node $i$ transmits to node $j$ using a specific channel, this transmission would be successfully received by node $j$, if
\small
\begin{equation} \label{eq:1} |X_{k}-X_{j}| \geq (1 + \Delta)|X_{i}-X_{j}|	\end{equation}
\normalsize

\noindent for every node $k$ simultaneously transmitting over the same channel. $\Delta$ is related to power margin required to ensure the successful reception at node $j$ even though node $k$ transmits at the same time. The Gupta and Kumar's Protocol Model is said to consider only the DATA to DATA collision constraints between two simultaneous transmitting links. Liew \cite{pcng} pointed though Gupta and Kumar's proposed model is named as a Protocol Model it does not fully characterize the medium access protocol being used. Hence, Liew proposed another model \cite{ivan} where Physical Collision Constraints and Protocol Collision Prevention Constraints among the DATA and protocol specific control packets were considered. Basel et al. \cite{alawieh2} have also proposed a model considering the protocol components of a transmission. They studied the relationship between tuning carrier sensing threshold and transmission power control for Basic Access Scheme and RTS/CTS Access Scheme. Although the control packets may slightly reduce the collision among contending hosts, their impact on the spatial reuse and the added overhead outweigh their benefits specifically when used at high rates. This comparative study has showed that the Basic Access Scheme always outperforms the RTS/CTS Access Scheme. Although Liew's and Basel's proposals including the recent works in \cite{tinnirello,fu} reflect a more accurate model as they have considered a protocol model, they are only suitable for network using OA. We model interference using protocol model for network using DA.

Li et al. \cite{li} have investigated the capacity of wireless networks using DAs. They proposed that the number of beams of DAs need to increase as the number of nodes increases in order for both random and arbitrary networks to scale. Although Li's proposal including the recent works in \cite{xu,suyi,alawieh1} have modeled interference for network using DA they have not proposed a metric to measure the severity of interference. In fact, there are not many works done to quantize the severity of interference in an aggregated form for a wireless network. Parameters such as throughput and packet error ratio do not directly explain the interference in a wireless network. SINR is perhaps the closest way to quantize interference, but it is not a global metric. Liew \cite{ivan} proposed the \textit{Attacking Case}, a metric that considers the interference caused by protocol dependent and protocol independent constraints which are captured in graphs. Although very good, the approach was developed for nodes using omnidirectional antenna. We extend the \textit{Attacking Case} metric to cater for nodes using DA.

\section{Power Constraints in IEEE 802.11 Network}

A node using DA is able to transmit at one specific angular direction at a time slot and later change direction to transmit at a different angle at another time slot.  In this section we extend the Physical Collision Constraints and Protocol Collision Prevention Constraints proposed in \cite{ivan} to accommodate DA. At the end of the section we discuss the differences between our proposed extensions and Liew's models.

\subsection{Physical Collision Constraints}
\begin{figure}
\centering
\includegraphics[clip,scale=1.3]{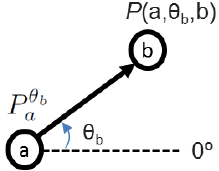}
\caption{Transmission power notation for Node $a$ transmitting to Node $b$}
\label{fig:pa}
\end{figure}

The Physical Collision Constraints can be modeled using the pair-wise interference model. For a link under the pair-wise interference model, the interferences from the other links are considered one by one. In particular, the pair-wise interference model does not take into account the cumulative effects of the interferences from the other links \cite{fu}.
\small
\begin{equation} \label{eq:2} P\left(\textit{a},\theta_{b},\textit{b}\right) = c\left(\textit{a},\theta_{b},\textit{b}\right) \cdot P^{\theta_{b}}_{a}/r^{\alpha} \end{equation}
\normalsize

\noindent where $P\left(\textit{a},\theta_{b},\textit{b}\right)$ is the power received by node $b$ from the direction $\theta_{b}$ of node $a$ and $P^{\theta_{b}}_{a}$ is the power transmitted by node $\textit{a}$ in the direction of node $b$ as shown in Fig. \ref{fig:pa}. $r$ is the distance between the two nodes, $\alpha$ is the path-loss exponent, and $c\left(\textit{a},\theta_{b},\textit{b}\right)$ is a constant in the direction of node $b$ from node $a$. For instance for two-ray ground reflection radio propagation model $\alpha$ is 4 and $c\left(\textit{a},\theta_{b},\textit{b}\right)$ is defined as in Eq. \ref{eq:22ray}.
\small
\begin{equation} \label{eq:22ray} c\left(\textit{a},\theta_{b},\textit{b}\right) = (G^{\theta_{b}}_{a} \cdot G^{((\theta_{b} + 180^{\circ})\ mod\ 360^{\circ})}_{b} \cdot h^{2}_{a} \cdot h^{2}_{b}) \end{equation}
\normalsize

\noindent where $G^{\theta_{b}}_{a}$ is the gain of node $a$'s antenna in the direction of node $b$ and $G^{((\theta_{b} + 180^{\circ})\ mod\ 360^{\circ})}_{b}$ is gain of node $b$'s antenna in the direction of node $a$. $h_{a}$ and $h_{b}$ are the heights of node $a$'s and node $b$'s antennas respectively. Similar relationship as in Eq. \ref{eq:22ray} can be derived for other radio propagation models. $\theta_{(.)}$ is suitable to represent any type of directional antenna such as adaptive array antenna, switched beam antenna or several elements of passive directional antennas connected via multiple interfaces. The present definition is straight forward for adaptive array antenna; in switched beam antenna $\theta_{(.)}$ will translate to the $beam\_id$ that radiates in the direction of angle $\theta_{(.)}$; in multi-interface directional antenna system $\theta_{(.)}$ will translate to the $interface\_id$ that radiates in the direction of angle $\theta_{(.)}$.

Let us consider two pairs of data links, Link $i$ and Link $j$, communicating using the Basic Access Scheme of IEEE 802.11 MAC protocol (DATA and ACK) without RTS and CTS. Let $T_{i}$ and $R_{i}$ represent respectively the position of the transmitter and receiver of Link $i$. $T_{i}$ and $R_{i}$ are also used for simplicity to refer to the nodes. $T_{i}$ will transmit DATA and receive ACK while $R_{i}$ will receive DATA and transmit ACK. Four different possible combination of simultaneous transmissions by Link $i$ and Link $j$ may occur: DATA$_{i}$-DATA$_{j}$, DATA$_{i}$-ACK$_{j}$, ACK$_{i}$-DATA$_{j}$, and ACK$_{i}$-ACK$_{j}$. The following Physical Collision Constraints can be derived for the four combinations. For a DATA$_{i}$-DATA$_{j}$ pair of transmissions a collision occurs when Link $i$ interferes with Link $j$. The transmission of Link $i$ will be interfering with the transmission of Link $j$ if,
\small
\begin{equation} \label{eq:3} 
P({T_{j}},\theta_{R_{j}},{R_{j}}) < K P({T_{i}},\theta_{R_{j}},{R_{j}}) \textnormal{\hspace{20pt}(DATA$_{i}$-DATA$_{j}$)}\\
\end{equation}
\normalsize

\noindent where $K$ is the Signal to Interference Ratio (SIR) requirement for a packet to be successfully decoded by the IEEE 802.11 protocol (e.g 10 dB). Independently of $T_{i}$ transmitting first or $T_{j}$ transmitting first, as long as the two transmissions overlap in time, $T_{j}$'s DATA transmission will be interfered at $R_{j}$ if the constraint in Eq. \ref{eq:3} is satisfied. Similar relationships can be established for the other 3 constraints. The transmission of Link $i$ will interfere with the transmission of Link $j$ if,
\small
\begin{gather} \label{eq:5}
P({R_{j}},\theta_{T_{j}},{T_{j}}) < K P({T_{i}},\theta_{T_{j}},{T_{j}}) \textnormal{\hspace{20pt}(DATA$_{i}$-ACK$_{j}$)}\\
P({T_{j}},\theta_{R_{j}},{R_{j}}) < K P({R_{i}},\theta_{R_{j}},{R_{j}}) \textnormal{\hspace{20pt}(ACK$_{i}$-DATA$_{j}$)}\\
P({R_{j}},\theta_{T_{j}},{T_{j}}) < K P({R_{i}},\theta_{T_{j}},{T_{j}}) \textnormal{\hspace{20pt}(ACK$_{i}$-ACK$_{j}$)}
\end{gather}
\normalsize

\subsection{Protocol Collision Prevention Constraints}
The Protocol Collision Prevention Constraints of IEEE 802.11 consider the effect of carrier sensing. The goal of carrier sensing is to prevent simultaneous transmissions. The prevention of a transmission can be induced at the transmitter's side, at the receiver's side or at both sides. There are two types of carrier sensing that would prevent a transmission: 

\textbf{Physical Carrier Sensing (PCS) -} The PCS defined by IEEE is the Clear Channel Assessment (CCA) mechanism \cite{mac}. When a carrier is sensed by the radio interface, the CCA mechanism indicates a busy medium and prevents the radio interface from initiating its own transmission. If a node is within the carrier sensing range (CSRange) of a transmitting node, in presence of no other interference, the PCS mechanism of the node would be triggered every time a packet is detected. 

\textbf{Virtual Carrier Sensing (VCS) -} The VCS mechanism is defined in addition to the PCS \cite{mac}. VCS uses the information found in IEEE 802.11 packets to predict the status of the wireless medium and determine how long a node has to wait before attempting to transmit. If a node is within the transmission range (TXRange) of a transmitting node, in presence of no other interference, the VCS mechanism of the node would be triggered every time a packet is being detected. 

\subsubsection{Transmitter Side}

A transmitter would refrain from transmitting a DATA packet if it can sense the transmission of another ongoing transmission. The transmission of Link $i$ will interfere with the transmission of Link $j$ if,
\small
\begin{gather} \label{eq:6}
|T_{j}-T_{i}| < CSRange(P^{\theta_{T_{j}}}_{T_{i}}) \textnormal{\hspace{34pt}(DATA$_{i}$-DATA$_{j}$)}\\
|T_{j}-R_{i}| < CSRange(P^{\theta_{T_{j}}}_{R_{i}}) \textnormal{\hspace{37pt}(ACK$_{i}$-DATA$_{j}$)}\\
|T_{j}-T_{i}| < TXRange(P^{\theta_{T_{j}}}_{T_{i}}) \textnormal{\hspace{28pt}(DATA$_{i}$-DATA$_{j}$)}
\end{gather}
\normalsize


\subsubsection{Receiver Side}
In IEEE 802.11 commercial products, when $T_{i}$ is already transmitting, $T_{j}$ can still transmit if $T_{i}$ interferes only with $R_{j}$ but not $T_{j}$. However, $R_{j}$ will ignore the DATA from $T_{j}$ and not transmit an ACK to $T_{j}$ fearing it may interfere with the ongoing transmission on Link $i$ \cite{ivan}. The transmission of Link $i$ will interfere with the transmission of Link $j$ if,
\small
\begin{gather} \label{eq:7}
|R_{j}-T_{i}| < CSRange(P^{\theta_{R_{j}}}_{T_{i}}) \textnormal{\hspace{33pt}(DATA$_{i}$-ACK$_{j}$)}\\
|R_{j}-R_{i}| < CSRange(P^{\theta_{R_{j}}}_{R_{i}}) \textnormal{\hspace{37pt}(ACK$_{i}$-ACK$_{j}$)}\\
\label{eq:7a}
|R_{j}-T_{i}| < TXRange(P^{\theta_{R_{j}}}_{T_{i}}) \textnormal{\hspace{31pt}(DATA$_{i}$-ACK$_{j}$)}
\end{gather}
\normalsize

\subsection{Power Constraints by Liew}

Liew, in \cite{ivan}, has modeled the Physical Collision Constraints using Eq. \ref{eq:8}. As we are modeling a network with nodes that use DA, Eq. 14 is not suitable for such a network. We have extended Eq. \ref{eq:8} by incorporating the direction of transmission $\theta$ as shown in Eq. 2. 
\small
\begin{equation} \label{eq:8} P\left(\textit{a},\textit{b}\right) = c \cdot P_{a}/r^{\alpha} \end{equation}
\normalsize

Liew has considered the Virtual Carrier Sensing Range \mbox{(VCSRange)} and the Physical Carrier Sensing Range \mbox{(PCSRange)} when modeling the Protocol Collision Prevention Constraints. \mbox{VCSRange} refers to the virtual carrier sensing ranges by the transmission of RTS/CTS packets and \mbox{PCSRange} refers to the physical carrier sensing ranges by the transmission of DATA packets \cite{ivan}. For the correct operation of the physical layer we have considered the CSRange and TXRange which is limited by the carrier sensing range and transmission ranges of any packets sent over a wireless channel. This is because non-RTS/CTS packets such as DATA do also have VCS functionally. 

\section{Graph Models for Attacking Case}
In this section the Physical Collision Constraints and the Protocol Collision Prevention Constraints are used to model 3 weighted directed graphs: the Link-Interference Graph, the Transmitter-side Protocol Collision Prevention Graph, and the Receiver-side Protocol Collision Prevention Graph. These graphs will be used to construct our improved $\textit{Attacking}$ $\textit{Case}$ metric. Let us define the general graph $G$ as a collection of vertices $V$ and unidirectional edges $E$ that connect pairs of vertices with weights $w$.  
\small
\begin{equation} \label{eq:9}
G = (V, E, w) 
\end{equation}
\normalsize

\noindent For any unidirectional edge $e_{ij} \in E$ where $i,j \in V$, vertex $i$ represents Link $i$ consisting of T$_{i}$ and R$_{i}$ nodes, while $e_{ij}$ represents a relationship between Link $i$ and Link $j$. The weight is a function of $e_{ij}$ where $w(e_{ij}) \in \mathds{N}$. The value of $w(e_{ij})$ depends on the type graph being modeled.

\begin{figure}
\centerline{
\subfloat[Network 1]{\includegraphics[clip,scale=1.4]{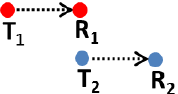}\hspace{3em}
\label{fig:case1}}
\subfloat[Network 2]{\includegraphics[clip,scale=1.4]{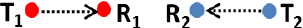}
\label{fig:case2}}}
\caption{Example networks - Network 1 and Network 2 - used to capture different interference conditions and to present the 3 graphs.}
\label{fig:Networks Setup}
\end{figure}

We introduce the 3 proposed graphs by discussing two simple networks: Network 1 and Network 2, shown in Fig. \ref{fig:Networks Setup}. The distances of transmitter-receiver pairs, $R_{1}$ and $T_{2}$ in Network 1, and $R_{1}$ and $R_{2}$ in Network 2 are 200 m. Each network is analyzed for 3 different setups where a setup is characterized by the type of antenna used (omnidirectional, directional) and by the ranges of a node (TXRange, CSRange). For the sake of analysis simplicity, ranges are defined based on a two-ray ground reflection radio propagation model and the effect of cross over distance and random component for shadowing are not considered. K is set to 10 dB.

The 3 setups addressed are the following:

\begin{enumerate}[(a)]
	\item Omnidirectional Antenna Setup (OA Setup) - Antenna= Omnidirectional, Gain= 1, Node's transmission power $P_{OA}$= 282 mW, TXRange= 250 m, CSRange= 550 m;
	\item Directional Antenna Setup (DA Setup) - Antenna= Directional ($90^{^{\circ}}$ beamwidth), Gain= 2, Node's transmission power $P_{DA}$ = $P_{OA}$, TXRange= 374 m, CSRange= 778 m; 
	\item Directional Antenna with Reduced Transmit Power Setup (DR Setup) - Antenna= Directional ($90^{^{\circ}}$ beamwidth), Gain= 2, Node's transmission power such that $transmit\ range\ $($P_{DR}$)\ = $transmit\ range\ $($P_{OA}$), TXRange= 250 m, CSRange= 550 m.
\end{enumerate}

\begin{figure}
\centerline{
\subfloat[Network 1 - OA Setup]{\includegraphics[clip,scale=0.8]{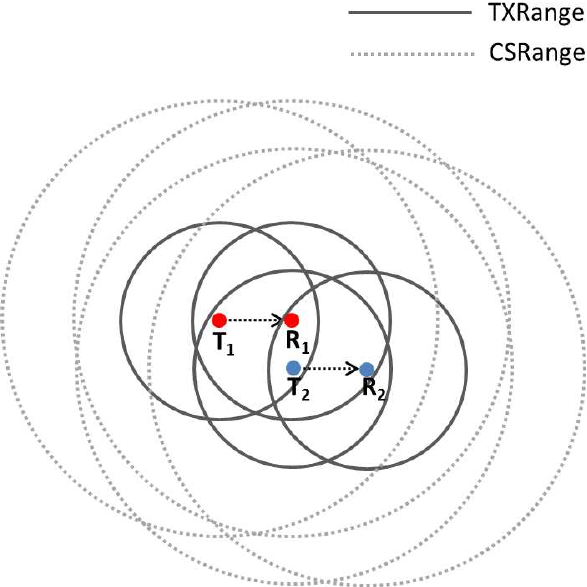}
\label{fig:a_oa}}
\subfloat[Network 2 - OA Setup]{\includegraphics[clip,scale=0.8]{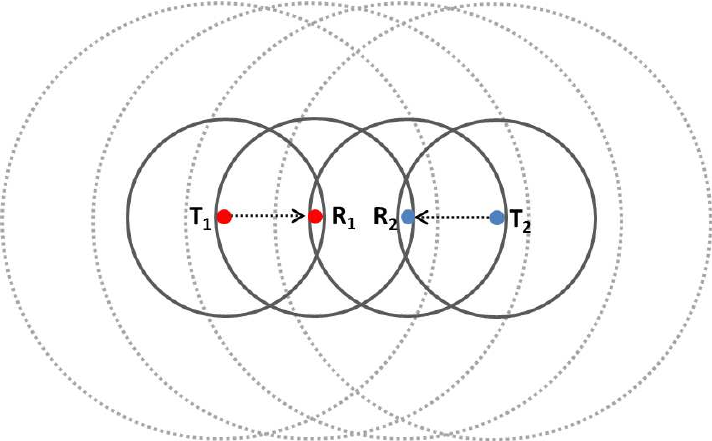}
\label{fig:b_oa}}}
\centerline{
\subfloat[Network 1 - DA Setup]{\includegraphics[clip,scale=0.8]{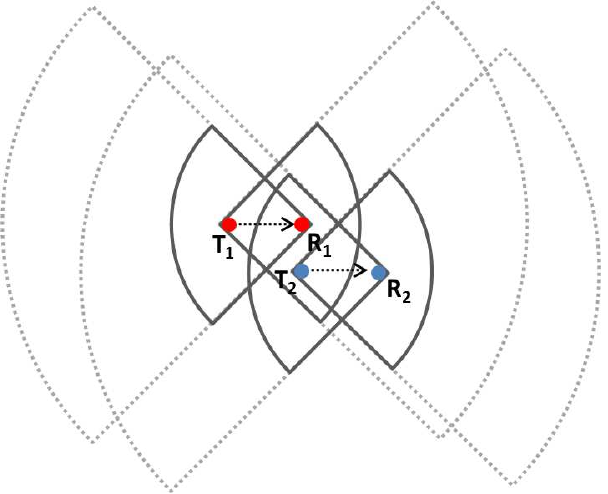}
\label{fig:a_da}}
\subfloat[Network 2 - DA Setup]{\includegraphics[clip,scale=0.8]{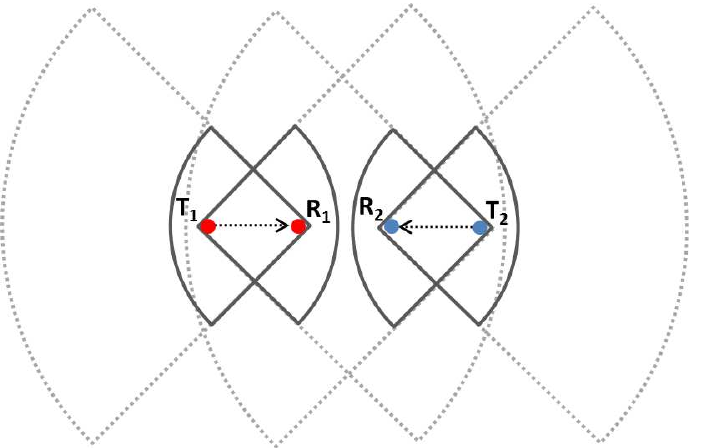}
\label{fig:b_da}}}
\centerline{
\captionsetup[subfigure]{margin={-1.2cm,0cm}}
\subfloat[Network 1 - DR Setup]{\includegraphics[clip,scale=0.8]{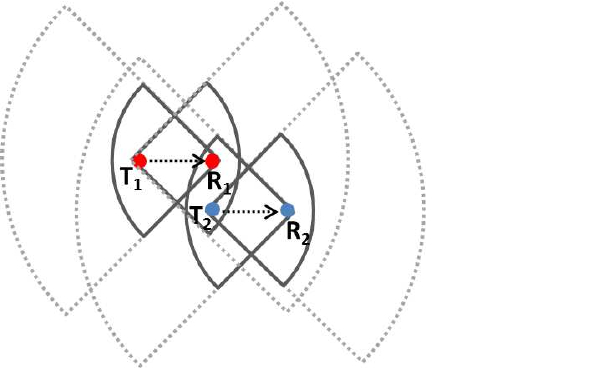}
\label{fig:a_dr}}
\subfloat[Network 2 - DR Setup\hspace*{-15 mm}]{\includegraphics[clip,scale=0.8]{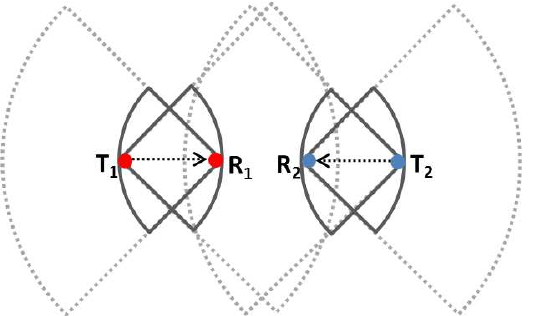}
\label{fig:b_dr}}}
\caption{TXRanges and CSRanges representation for 3 setups for Network 1 and Network 2}
\label{fig:ab}
\end{figure}

Fig. \ref{fig:ab} describes the 2 networks and the 3 setups along with their TXRanges and CSRanges. 

\subsection{Link-Interference Graph (i-graph)}
A Link-Interference Graph is used to represent the Physical Collision Constraints and it captures the SIR effects among links. The graph is represented as follows:
\small
\begin{equation} \label{eq:10}
G_{I} = (V_{I}, E_{I}, w_{I}) 
\end{equation}
\normalsize

\begin{figure}
\centering
\includegraphics[clip,scale=1.5]{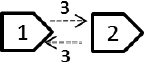}
\caption{i-graph for the network in Fig. \ref{fig:a_oa}}
\label{fig:mingraph}
\end{figure}

The i-graph of the network topology illustrated in Fig. \ref{fig:a_oa} can be represented by the graph in Fig. \ref{fig:mingraph}. In the figure, an arrow-shaped vertex represents a wireless link with the arrow pointing towards the receiver of the link. Each vertex is labeled with the $link\_id$ (Link 1 or Link 2) it represents. An arrow connects vertex 1 to vertex 2 if there is a relationship from Link 1 to Link 2. The edge $e_{ij}$ is labeled with its $w_{I}(e_{ij})$. 

Consider the topology of Fig. \ref{fig:a_oa} where the nodes use OA. There is a directional i-edge, shown in Fig. \ref{fig:mingraph}, from vertex 2 to vertex 1 because the transmitter of Link 2 interferes with receiver of Link 1. More specifically, DATA transmitted by $T_{2}$ will collide with a DATA transmitted by $T_{1}$ at $R_{1}$ if the transmissions overlap in time since, in this case, Eq. \ref{eq:3} holds (DATA$_{2}$-DATA$_{1}$). In the reverse direction, there is no i-edge from vertex 1 to vertex 2 due to DATA$_{1}$-DATA$_{2}$ pair of transmission but there is an i-edge from vertex 1 to vertex 2 due to DATA$_{1}$-ACK$_{2}$, ACK$_{1}$-DATA$_{2}$, and ACK$_{1}$-ACK$_{2}$ pairs of transmission. There are also i-edges from vertex 2 to vertex 1 due to DATA$_{2}$-ACK$_{1}$ and ACK$_{2}$-DATA$_{1}$ pairs of transmissions.

In general, if any of the constraints in Eq. 4, 5, 6 or 7 is satisfied, an edge would be drawn from vertex $i$ to vertex $j$ to signify that Link $i$ is interfering with Link $j$. We propose that the unidirectional edge in the i-graph has a weight $w_{I}(e_{ij})$ characterized as follows:
\small
\begin{equation} \label{eq:11}
\begin{split}
w_{I}(e_{ij}) = &\mathds{1}_{\left[P^{\theta_{R_{j}}}_{T_{j}}|T_{i}-R_{j}|^{\alpha} < KP^{\theta_{R_{j}}}_{T_{i}}|T_{j}-R_{j}|^{\alpha}\right]} + \\
            &\mathds{1}_{\left[P^{\theta_{T_{j}}}_{R_{j}}|T_{i}-T_{j}|^{\alpha} < KP^{\theta_{T_{j}}}_{T_{i}}|T_{j}-R_{j}|^{\alpha}\right]} + \\
						&\mathds{1}_{\left[P^{\theta_{R_{j}}}_{T_{j}}|R_{i}-R_{j}|^{\alpha} < KP^{\theta_{R_{j}}}_{R_{i}}|T_{j}-R_{j}|^{\alpha}\right]} + \\
						&\mathds{1}_{\left[P^{\theta_{T_{j}}}_{R_{j}}|R_{i}-T_{j}|^{\alpha} < KP^{\theta_{T_{j}}}_{R_{i}}|T_{j}-R_{j}|^{\alpha}\right]} 
\end{split}						
\end{equation}
\normalsize

\noindent where Eq. \ref{eq:11} is built using components of characteristic function as defined in Eq. \ref{eq:12}.
\small
\begin{equation} \label{eq:12}
   \mathds{1}_{[C]} = \left\{
     \begin{array}{lr}
       1, \ \ if \;C = TRUE\\
       0, \ \ if \;C = FALSE
     \end{array}
   \right.			
\end{equation}
\normalsize
Since $w_{I}(e_{ij})$ exists only when there is an $e_{ij}$, $w_{I}(e_{ij}) \in \left\{1,2,3,4\right\}$ for i-graph. For the OA setup in Fig. \ref{fig:a_oa}, its i-graph has directional edge from vertex 1 and vertex 2 and vice versa with weight $w_{I}(e_{12})$ = $w_{I}(e_{21})$ = 3.

\begin{figure}
\centering
\includegraphics[clip,width=1\columnwidth]{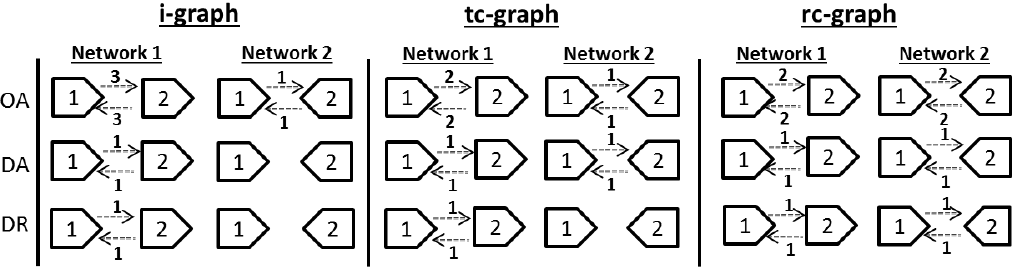}
\caption{Graph Models of the networks and setups presented in Fig. \ref{fig:ab} using our proposed method}
\label{fig:Graph SK Liew}
\end{figure}

In Fig. \ref{fig:a_da} the antenna is directional. Although i-edges exist as in OA setup from vertex 1 to vertex 2 due to DATA$_{1}$-ACK$_{2}$ pair of transmissions and vice versa, the i-edges due to the other transmission pairs do not exist. The ability of DA to point its beam to its intended destination reduces interference on unwanted directions. For the setup in Fig. \ref{fig:a_da}, $w_{I}(e_{12})$ = $w_{I}(e_{21})$ = 1 and the i-graph obtained can be observed in Fig. \ref{fig:Graph SK Liew}.

In Fig. \ref{fig:a_dr} the i-graph obtained is the same as in DA setup, where i-edges exist from vertex 1 to vertex 2 due to DATA$_{1}$-ACK$_{2}$ pair of transmissions and vice versa. The reduction of transmission power has no gain for i-graph in this topology. For the setup in Fig. \ref{fig:a_dr}, $w_{I}(e_{12})$ = $w_{I}(e_{21})$ = 1 and Fig. \ref{fig:Graph SK Liew} shows the i-graph obtained. 

In Fig. \ref{fig:b_oa} a different node positioning is tested and the nodes use OA. In the figure we can observe that there are directional i-edges from vertex 1 to vertex 2 due to ACK$_{1}$-DATA$_{2}$ pair of transmission and from vertex 2 to vertex 1 due to ACK$_{2}$-DATA$_{1}$ pair of transmission. $w_{I}(e_{12})$ = $w_{I}(e_{21})$ = 1 for the i-graph and this is shown in Fig. \ref{fig:Graph SK Liew}. We recall that in Fig. \ref{fig:a_oa} the weight was 3, hence the topology of a network affects the outcome of an i-graph and its edge's weight. 

In Fig. \ref{fig:b_da} and Fig. \ref{fig:b_dr} no pair of transmission creates an i-edge between vertex 1 and vertex 2, and vice versa; in these setups the antenna type plays an important role in eliminating edges between the vertices.

From Fig. \ref{fig:ab} and Fig. \ref{fig:Graph SK Liew} we can conclude that the DA and DR setups lead to the smallest interference. The OA setup has the highest value of weight on the i-edges. The more weight an i-edge has the more prone it gets for packet collision. Network 1 and Network 2 enable us to conclude that the topology affects the weight of an i-edge.
 
\subsection{Transmitter-side Protocol Collision Prevention Graph (tc-graph)}

Let us consider the effect of IEEE 802.11 carrier sensing. The goal of carrier sensing is to prevent simultaneous transmissions that will collide. The tc-graph models the effect of carrier sensing by the transmitters and it is represented as follows:
\small
\begin{equation} \label{eq:13}
G_{TC} = (V_{TC}, E_{TC}, w_{TC}) 
\end{equation}
\normalsize
 
In the tc-graph there is a directional tc-edge from vertex $i$ to vertex $j$ if $T_{j}$ can sense the transmission on Link $i$ so that, if $T_{i}$ or $R_{i}$ are already transmitting respectively a DATA or ACK packet, $T_{j}$ will not transmit. Formally, there is a tc-edge from vertex $i$ to vertex $j$ if any of the Eq. 8, 9 or 10 holds true.

In Fig. \ref{fig:a_oa}, $T_{1}$ and $T_{2}$ are not sufficiently far apart and they can sense each other. There is directional tc-edge from vertex 1 to vertex 2 because the transmitter of Link 1 interferes with the transmitter of Link 2. Specifically, the transmission of DATA from $T_{1}$ and ACK from $R_{1}$ will prevent DATA from $T_{2}$ to be transmitted. There is also a directional tc-edge in the reverse direction; the transmission of DATA from $T_{2}$ and ACK from $R_{2}$ will prevent DATA from $T_{1}$ for being transmitted. 

The edge in the tc-graph has a weight $w_{TC}(e_{ij})$ characterized as follows:
\small
\begin{equation} \label{eq:14}
\begin{split}
w_{TC}(e_{ij}) = &\mathds{1}_{\left[(|T_{j}-T_{i}| < CSRange(P^{\theta_{T_{j}}}_{T_{i}})) \vee (|T_{j}-T_{i}| < TXRange(P^{\theta_{T_{j}}}_{T_{i}}))\right]} + \\
            &\mathds{1}_{\left[|T_{j}-R_{i}| < CSRange(P^{\theta_{T_{j}}}_{R_{i}})\right]}	
\end{split}
\end{equation}
\normalsize

Since $w_{TC}(e_{ij})$ exists only when there is an $e_{ij}$, $w_{TC}(e_{ij}) \in \left\{1,2\right\}$ for tc-graph. For the setup in Fig. \ref{fig:a_oa}, $w_{TC}(e_{12})$ = $w_{TC}(e_{21})$ = 2 and the tc-graph obtained can be observed in Fig. \ref{fig:Graph SK Liew}. 

As the tc-graph models the effect of carrier sensing purely from the transmitter point of view, it does not consider tc-edges created due to the DATA$_{1}$-ACK$_{2}$ and ACK$_{1}$-ACK$_{2}$ pairs of transmission from vertex 1 to vertex 2 and DATA$_{2}$-ACK$_{1}$ and ACK$_{2}$-ACK$_{1}$ pairs of transmission from vertex 2 to vertex 1 due to its effect solely at the receiver.

In Fig. \ref{fig:a_da} the antenna is directional. There are tc-edges from vertex 1 to vertex 2 due to DATA$_{1}$-DATA$_{2}$ pair of transmission and from vertex 2 to vertex 1 due to ACK$_{2}$-DATA$_{1}$ pair of transmission. The tc-edges which occur in OA setup for ACK$_{1}$-DATA$_{2}$ and DATA$_{2}$-DATA$_{1}$ do not exist in DA setup. This is because of the ability of DA to point its beam to its intended receiver which also reduces interference to unwanted directions. For the setup in Fig. \ref{fig:a_da}, $w_{TC}(e_{12})$ = $w_{TC}(e_{21})$ = 1 and its tc-graph is shown in Fig. \ref{fig:Graph SK Liew}.

In Fig. \ref{fig:a_dr} the tc-graph is the same as for the DA setup, where tc-edges exist from vertex 1 to vertex 2 due to DATA$_{1}$-DATA$_{2}$ and from vertex 2 to vertex 1 due to ACK$_{2}$-DATA$_{1}$ pairs of transmission. As in i-graph, the transmission power reduction has no gain for tc-graph for this topology. $w_{TC}(e_{12})$ = $w_{TC}(e_{21})$ = 1 for the scheme in Fig. \ref{fig:a_dr}, and Fig. \ref{fig:Graph SK Liew} shows the tc-graph observed. 

For Network 2 using OA (Fig. \ref{fig:b_oa}) there are directional tc-edges from vertex 1 to vertex 2 due to ACK$_{1}$-DATA$_{2}$ pair of transmission and from vertex 2 to vertex 1 due to ACK$_{2}$-DATA$_{1}$ pair of transmission. The weight, $w_{TC}(e_{12})$ = $w_{TC}(e_{21})$ = 1. We recall that in Fig. \ref{fig:a_oa} the weight was 3 and reaffirm that network topology affects the outcome of an tc-graph and its edge's weight. 

In Fig. \ref{fig:b_da} the antenna is directional. The ACK$_{1}$-DATA$_{2}$ and ACK$_{2}$-DATA$_{1}$ pairs of transmission which were present in the OA setup do not cause tc-edges anymore, but the DATA$_{1}$-DATA$_{2}$ and vice versa pairs of transmission cause tc-edges for the DA setup. This is because though interference is able to be contained on unwanted direction, it actually increased in the direction of transmission when DA is used. For the setup in Fig. \ref{fig:b_da}, $w_{TC}(e_{12})$ = $w_{TC}(e_{21})$ = 1 and its resultant tc-graph is shown in Fig. \ref{fig:Graph SK Liew}.

In Fig. \ref{fig:b_dr} none of the pairs of transmission create a tc-edge between vertex 1 and vertex 2 and vice versa. In this case, DA and transmission power reduction have played an important role in eliminating edges between the vertices.

From Fig. \ref{fig:ab} and Fig. \ref{fig:Graph SK Liew} we can conclude that the DA and DR setups lead to the smallest interference. The more weight a tc-edge has the more a node will trigger its exponential backoff mechanism. Network 1 and Network 2 enable us to conclude that, as in i-graph, the topology affects the weight of tc-edges.

\subsection{Receiver-side Protocol Collision Prevention Graph (rc-graph)}

In rc-graph the effect of carrier sensing by receivers is modeled. The graph is represented as follows:
\small
\begin{equation} \label{eq:15}
G_{RC} = (V_{RC}, E_{RC}, w_{RC}) 
\end{equation}
\normalsize

There is a directional rc-edge from vertex $i$ to vertex $j$ if $R_{j}$ can sense the transmission on Link $i$. Specifically, there is an rc-edge from vertex $i$ to vertex $j$ if any of Eq. 11, 12 or 13 is true. In the default mode of IEEE 802.11 commercial products, when $T_{i}$ is already transmitting, $T_{j}$ can still transmit if there is an rc-edge, but no tc-edge, from vertex $i$ to vertex $j$. However, $R_{j}$ will ignore the DATA frame and will not return an ACK \cite{ivan}. The rationale for $R_{j}$ not returning an ACK to $T_{j}$ is that the ACK may interfere with the ongoing transmission on Link $i$. 

In Fig. \ref{fig:a_oa}, $R_{1}$ and $R_{2}$ are so close to each other that the DATA and ACK transmission of Link 1 can be sensed by $R_{2}$ and the DATA and ACK transmission of Link 2 can be sensed by $R_{1}$. Thus, there is a directional rc-edge from vertex 1 to vertex 2 and vice versa. 

An edge in the rc-graph has a weight $w_{RC}(e_{ij})$ characterized as follows:
\small
\begin{equation} \label{eq:16}
\begin{split}
w_{RC}(e_{ij}) = &\mathds{1}_{\left[(|R_{j}-T_{i}| < CSRange(P^{\theta_{R_{j}}}_{T_{i}})) \vee (|R_{j}-T_{i}| < TXRange(P^{\theta_{R_{j}}}_{T_{i}}))\right]} + \\
            &\mathds{1}_{\left[|R_{j}-R_{i}| < CSRange(P^{\theta_{R_{j}}}_{R_{i}})\right]} 
\end{split}						
\end{equation}
\normalsize

Since $w_{RC}(e_{ij})$ exist only when there is an $e_{ij}$, $w_{RC}(e_{ij}) \in \left\{1,2\right\}$ for rc-graph. For the case of Fig. \ref{fig:a_oa}, $w_{RC}(e_{12})$ = $w_{RC}(e_{21})$ = 2 and its rc-graph is shown in Fig. \ref{fig:Graph SK Liew}.

Since rc-graph models the effect of carrier sensing purely from the receiver point of view, it does not consider rc-edges created due to the ACK$_{1}$-DATA$_{2}$ and DATA$_{1}$-DATA$_{2}$ pairs of transmission from vertex 1 to vertex 2, and ACK$_{2}$-DATA$_{1}$ and DATA$_{2}$-DATA$_{1}$ pairs of transmission from vertex 2 to vertex 1.

In Fig. \ref{fig:a_da} and in Fig. \ref{fig:a_dr} rc-edges were created in both the setups due to DATA$_{1}$-ACK$_{2}$ pair of transmission from vertex 1 to vertex 2 and ACK$_{2}$-ACK$_{1}$ pair of transmission from vertex 2 to vertex 1. For the cases of Fig. \ref{fig:a_da} and Fig. \ref{fig:a_dr}, $w_{RC}(e_{12})$ = $w_{RC}(e_{21})$ = 1 and its rc-graphs are shown in Fig. \ref{fig:Graph SK Liew}. DA has contributed to reduce the weight of the edges.

In Fig. \ref{fig:b_oa} there is rc-edge from vertex 1 to vertex 2 due to DATA$_{1}$-ACK$_{2}$ and ACK$_{1}$-ACK$_{2}$ pairs of transmission. There is also rc-edge from vertex 2 to vertex 1 due to DATA$_{2}$-ACK$_{1}$ and ACK$_{2}$-ACK$_{1}$ pairs of transmission. For the setup in Fig. \ref{fig:b_oa}, $w_{RC}(e_{12})$ = $w_{RC}(e_{21})$ = 2 and its resultant rc-graph is shown in Fig. \ref{fig:Graph SK Liew}.

In Fig. \ref{fig:b_da} and Fig. \ref{fig:b_dr} both the setups have rc-edges due to DATA$_{1}$-ACK$_{2}$ pair of transmission from vertex 1 to vertex 2 and DATA$_{2}$-ACK$_{1}$ pair of transmission from vertex 2 to vertex 1. The weight $w_{RC}(e_{12})$ = $w_{RC}(e_{21})$ = 1. 

From Fig. \ref{fig:ab} and Fig. \ref{fig:Graph SK Liew} we can conclude that the DA and DR setups are able to contain interference and assist in reducing the weight of the edges. The transmission power control has no advantage for these networks as the power reduced is still insufficient to curtail interference in the direction of DA's transmission. 

For i-graph, tc-graph and rc-graph all the vertices are the same, where $V$ = $V_{I}$ = $V_{TC}$ = $V_{RC}$. 

\subsection{Improved Attacking Case Metric}
$\textit{Attacking}$ $\textit{Case}$ corresponds to the number of cases where simultaneous transmissions are either not allowed or if allowed will not be successful. $\textit{Attacking}$ $\textit{Case}$ can be used as a performance metric to quantize the interference of a network. A high $\textit{Attacking}$ $\textit{Case}$ value leads to potentially poor aggregated network throughputs. We propose the following: 1) if $e_{i,j}$ is an i-edge then twice the i-edge's weight is added to the $\textit{Attacking}$ $\textit{Case}$ else; 2) if $e_{i,j}$ is a tc-edge then the tc-edge's weight is added to the $\textit{Attacking}$ $\textit{Case}$, and 3) if $e_{i,j}$ is a rc-edge then the rc-edge's weight is added to the $\textit{Attacking}$ $\textit{Case}$ for all $i$,$j$ where $i \neq j$ as shown in Eq. \ref{eq:17}.
\small
\begin{equation} \label{eq:17} 
\begin{split} 
\mathit{AC_{Imp}} = \smash[b]{\sum_{\substack{i,j \in \mathit{V}\\i\neq j}}}
\;\bigl[
  & 2 \times w_{I}(e_{i,j}) \times \mathds{1}_{[e_{i,j} \in E_{I}]} +{} \\
  & w_{TC}(e_{i,j}) \times \mathds{1}_{[e_{i,j} \in E_{TC} \wedge e_{i,j} \notin E_{I}]} +{} \\
  & w_{RC}(e_{i,j}) \times \mathds{1}_{[e_{i,j} \in E_{RC} \wedge e_{i,j} \notin E_{I}]}	
\bigr]
\end{split} 
\end{equation} 
\normalsize

Eq. \ref{eq:17} takes into account the order of transmissions. If $e_{i,j}$ is an i-edge, it does not matter whether Link $i$ or Link $j$ transmits first, the packet at Link $j$ will be corrupted. Hence, there are two cases where Link $i$ can interference with Link $j$. On the other hand if $e_{i,j}$ is a tc-edge or rc-edge, transmission at Link $j$ will not be allowed or will fail only if Link $i$ transmits first. So, there is only one case considered. 

\subsection{Graph Models for Liew's Attacking Case}

Liew in \cite{ivan} has modeled the $\textit{Attacking}$ $\textit{Case}$ using the graph model in Eq. \ref{eq:18}. 
\small
\begin{equation} \label{eq:18}
G = (V, E) 
\end{equation}
\normalsize

We have extended Eq. \ref{eq:18} by associating it with weights $w$ to the edge of the Link-Interference Graph, Transmitter-side Protocol Collision Prevention Graph, and Receiver-side Protocol Collision Prevention Graph as shown in Eq. \ref{eq:9}.

In Liew's method, if $e_{i,j}$ is an i-edge then 2 is added to the $\textit{Attacking}$ $\textit{Case}$, else if $e_{i,j}$ is a tc-edge then 1 is added to the $\textit{Attacking}$ $\textit{Case}$, else if $e_{i,j}$ is a rc-edge then 1 is added to the $\textit{Attacking}$ $\textit{Case}$ for all $i$,$j$ where $i \neq j$, as shown in Eq. \ref{eq:19}. We have improved Liew's method by considering the weights of the graphs and the method used to calculate the $\textit{Attacking}$ $\textit{Case}$ metric using the i-graph, tc-graph and rc-graph, as shown in Eq. \ref{eq:17}.
\small
\begin{equation} \label{eq:19} 
\begin{split} 
\mathit{AC_{Liew}} = \smash[b]{\sum_{\substack{i,j \in \mathit{V}\\i\neq j}}}
\;\bigl[
  & 2 \times \mathds{1}_{[e_{i,j} \in E_{I}]} +{} \\
	& \mathds{1}_{[e_{i,j} \in E_{TC} \wedge e_{i,j} \notin E_{I}]} +{} \\
  & \mathds{1}_{[e_{i,j} \in E_{RC} \wedge e_{i,j} \notin E_{TC} \wedge e_{i,j} \notin E_{I}]}	
\bigr]
\end{split} 
\end{equation} 
\normalsize

\section{Attacking Case Metric Evaluation}
In this section the improved $\textit{Attacking}$ $\textit{Case}$ metric (Eq. \ref{eq:17}) is used to quantize the severity of interference in CSMA/CA based networks by means of Network Simulator 2 (ns-2) simulations \cite{ns-2}. Firstly we show that the Liew's $\textit{Attacking}$ $\textit{Case}$ metric does not address nodes using DA. Secondly we show that our improved $\textit{Attacking}$ $\textit{Case}$ supports nodes using DA and it is also compatible for nodes using OA. Thirdly we show that our improved $\textit{Attacking}$ $\textit{Case}$ metric is able to quantize the interference for networks that use various transmission power schemes.

\subsection{ns-2 Simulator Enhancements}
\begin{figure}
\centering
\includegraphics[clip,width=0.5\columnwidth]{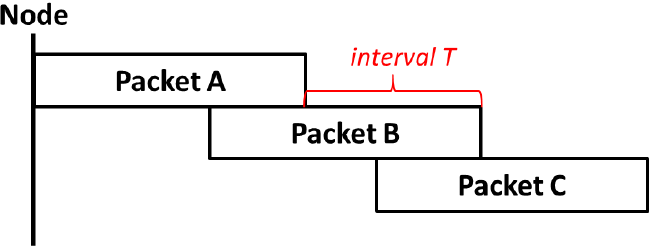}
\caption{Time interval T when packets A, B and C arrive at a Node}
\label{fig:periodx}
\end{figure}

When a node hears the arrival of packet A via CCA and if the received power is above a certain threshold, the packet is received by the node. First, the node's physical layer decodes the packet's Physical Layer Convergence Protocol (PLCP) Preamble and PLCP Header. In this process, the node will learn the characteristics of the forthcoming PLCP Service Data Unit (PSDU) such as the modulation used and length of the forthcoming PSDU segment in microseconds. Then the PHY-RXSTART primitive will be initiated if the cyclic redundancy check (CRC) of the PLCP header is positive. The length field of the PLCP header will determine the end of sending the PSDU octets to the MAC layer. This is done via the PHY-RXEND primitive. During the process of receiving packet A, if another packet B reaches this node overlapping in time and if its power is high enough, then the bits received from packet A are corrupted. The CRC check of packet A's PSDU will fail at the end of PHY-RXEND at the MAC layer. If any other packet, say packet C, reaches this node during the time interval $T$ of Fig. \ref{fig:periodx}, packet C may be received provided its received power is above the predefined SIR. The current behavior of ns-2 does not consider this aspect and disregards packet C \cite{ns-2}. We have extended the ns-2 simulator to consider this as we are studying scenarios operating in the overloaded conditions. 

\begin{figure}
\centering
\includegraphics[clip,width=0.8\columnwidth]{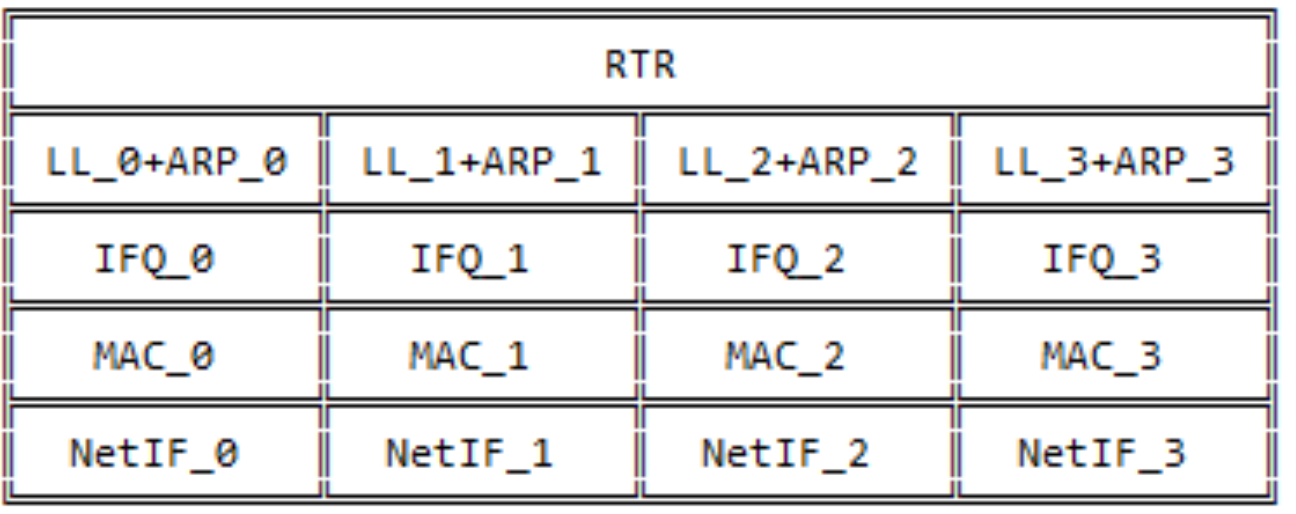}
\caption{Directional antenna stack for a wireless node in ns-2}
\label{fig:stackda}
\end{figure}

\begin{figure}
\centering
\includegraphics[clip,width=0.5\columnwidth]{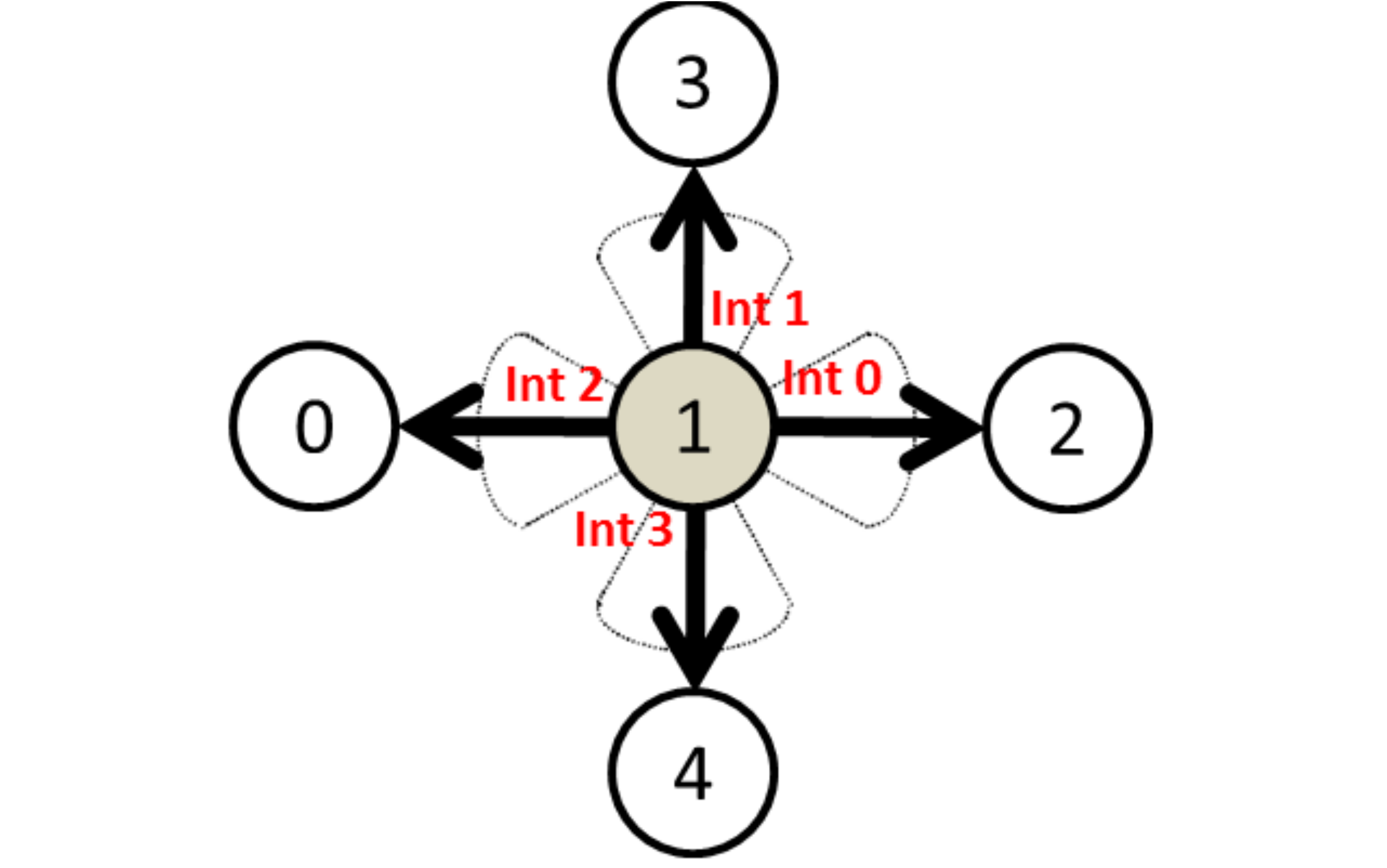}
\caption{Directional antenna model for a wireless node in ns-2}
\label{fig:da}
\end{figure}

ns-2 was also improved to support nodes with DA. Each node is assumed to have 4 interfaces where each interface is connected with an element of $90^{\circ}$ passive DA with ideal pie-slice radiation pattern of gain 2 without side or back lobe. The stack to support DA on a node is shown in Fig. \ref{fig:stackda} where each interface has a MAC, NAV, its own interface queue (IFQ), and maintains its own ARP table. The DA in interfaces 0, 1, 2 and 3 are pointed respectively to angle 0$^{\circ}$, 90$^{\circ}$, 180$^{\circ}$, and 270$^{\circ}$. As an example please refer to Node 1 in Fig. \ref{fig:da}. Node 1 reaches: Node 2 via Interface 0 pointed at $0^{\circ}$ angle; Node 3 via Interface 1 pointed at $90^{\circ}$ angle; Node 0 via Interface 2 pointed at $180^{\circ}$ angle; Node 4 via Interface 3 pointed at $270^{\circ}$ angle.

\subsection{Simulation Setup}
\begin{figure*}[ht!]
\centerline{{\includegraphics[clip,scale=0.5]{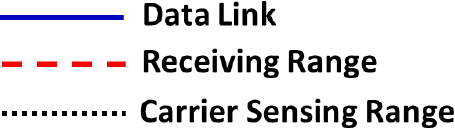}}}
\centerline{
\subfloat[Seed 14]{\includegraphics[clip,scale=0.2]{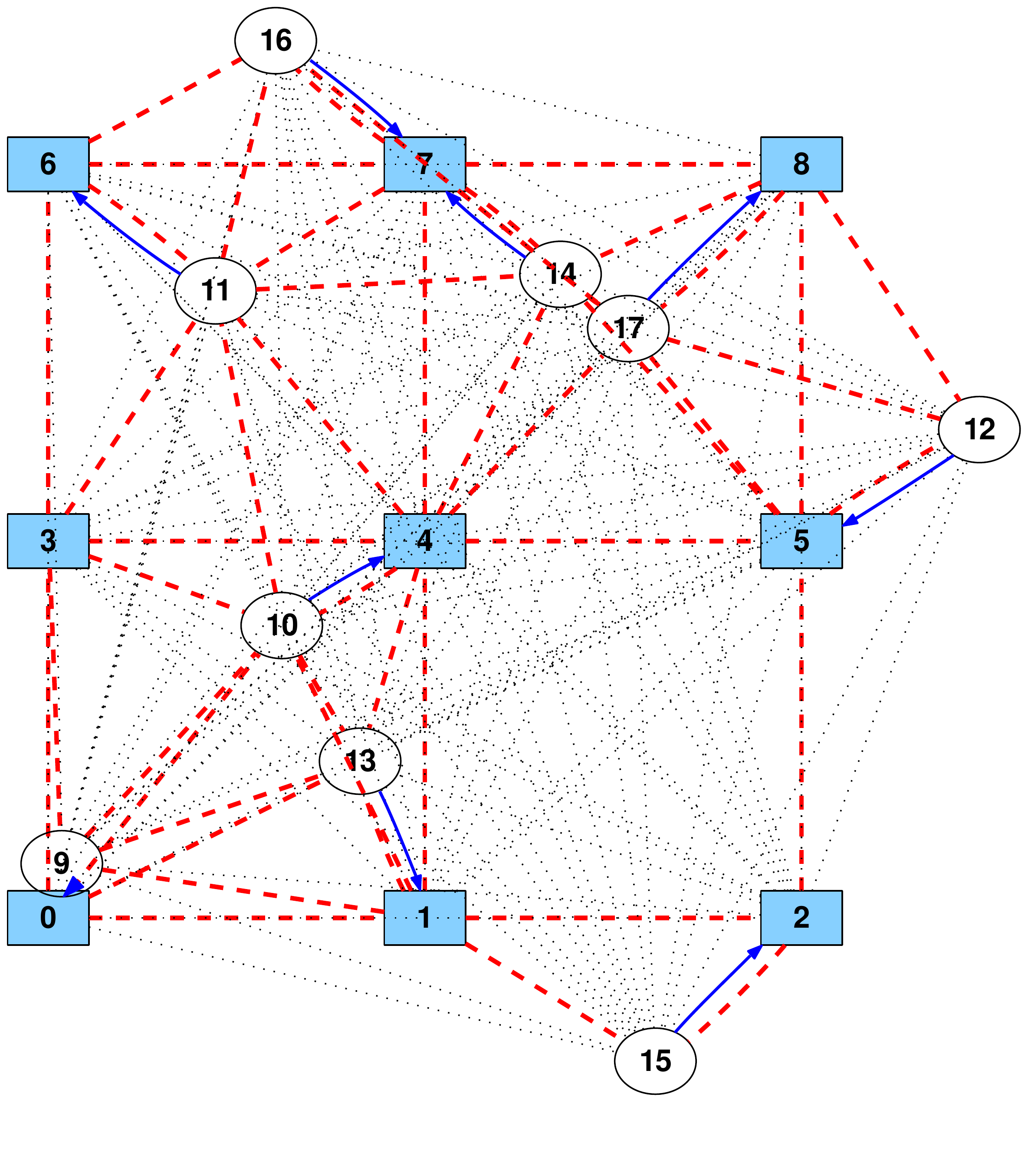}
\label{fig:topo14}}
\subfloat[Seed 21]{\includegraphics[clip,scale=0.2]{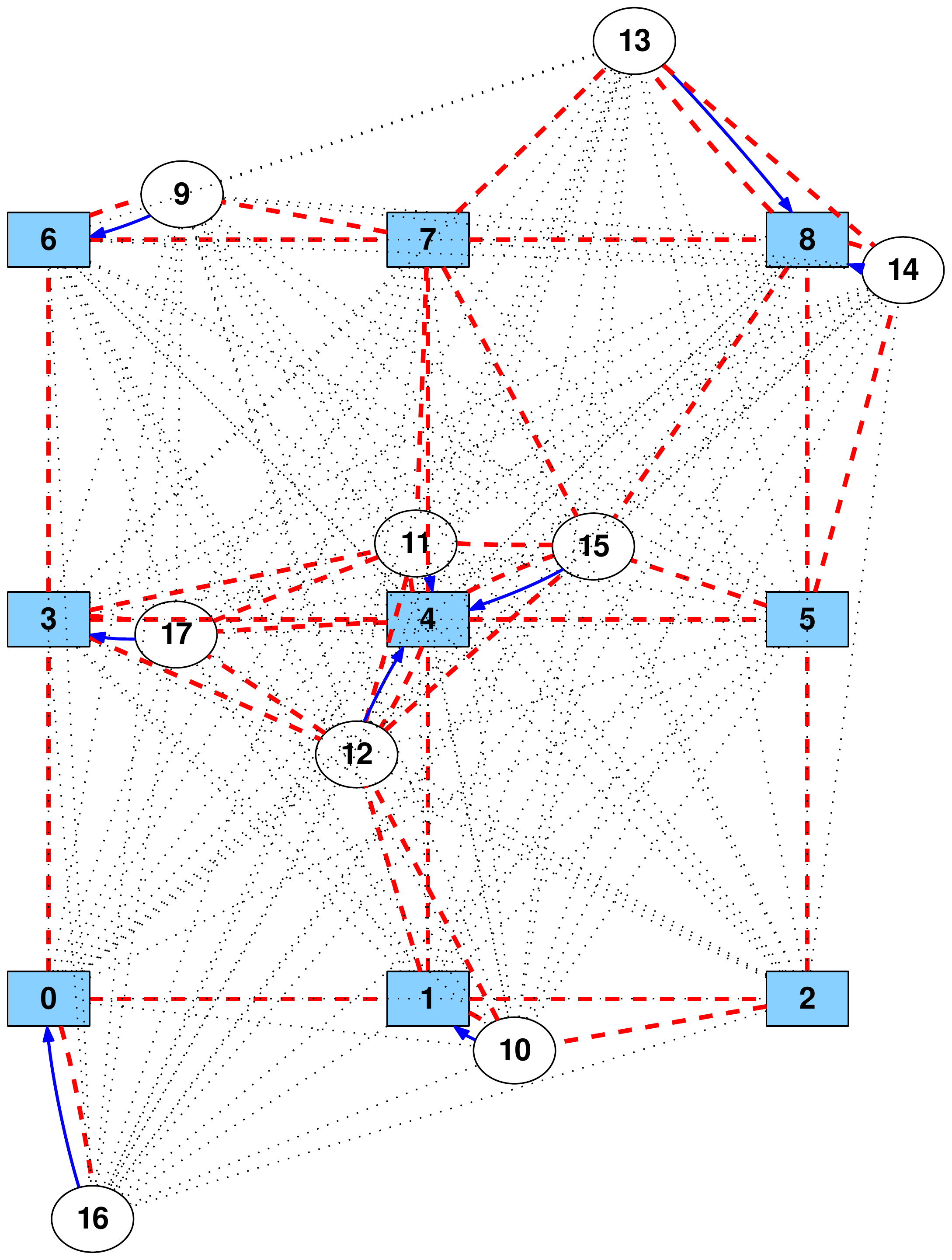}
\label{fig:topo21}}
\subfloat[Seed 100]{\includegraphics[clip,scale=0.2]{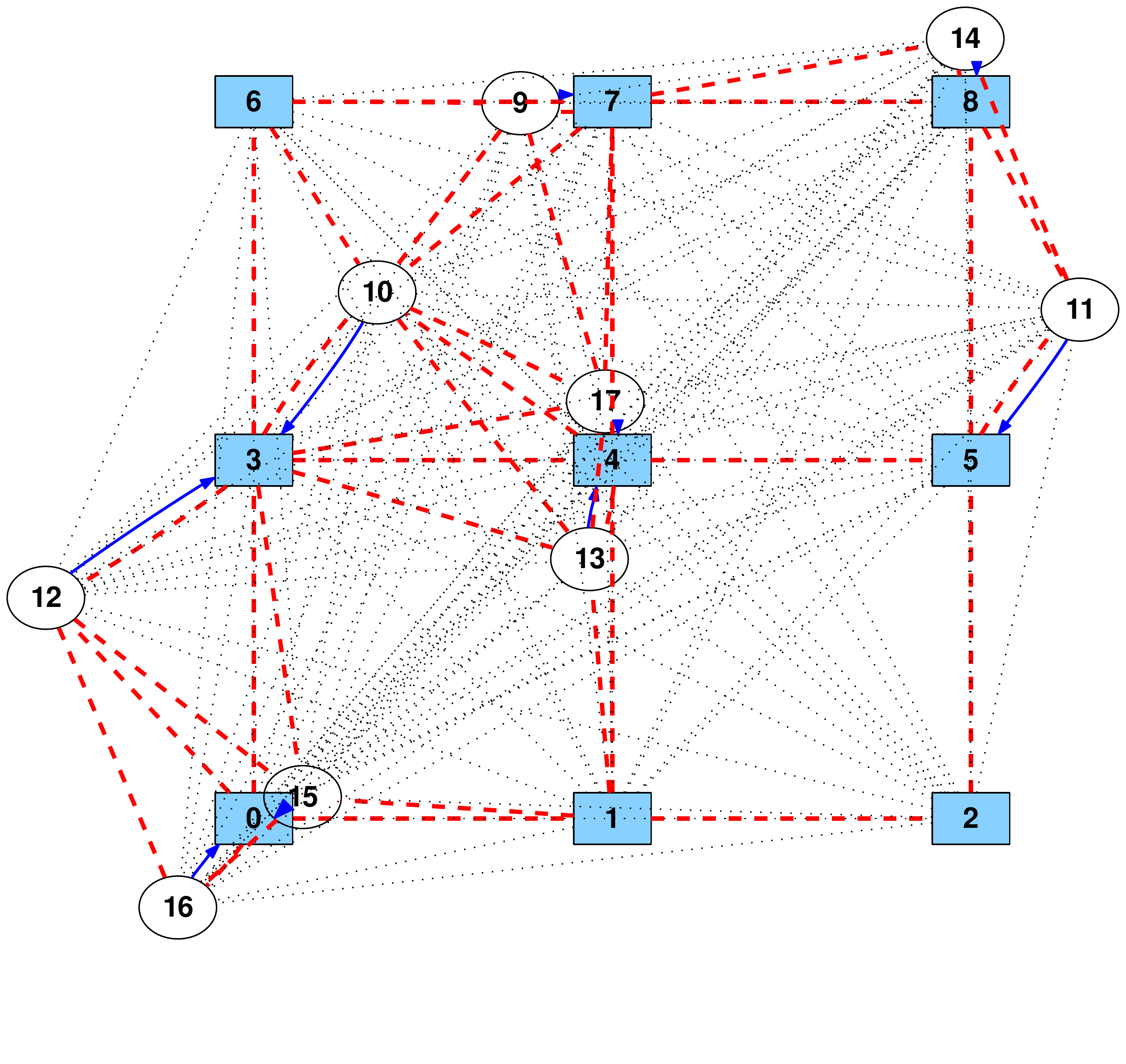}
\label{fig:topo100}}}
\caption{Example of random topologies for network with nodes using OA and 9 STAs}
\label{fig:toposeed}
\end{figure*}

We defined a 3 x 3 grid topology with nodes separated by 250 m and acting as APs. Additional nodes were placed randomly to represent STAs, where each STA will connect to the AP with the strongest signal which is naturally the closest AP. Traffic is sent from the STAs towards the APs replicating the video surveillance network scenario of \mbox{Fig. \ref{fig:topology}}. Being a single hop wireless network, routing was not considered. All the nodes are static. The number of random STAs in the network varied from 9 to 18, 27, and 36, aiming to increase the amount of interference in the network. For each scenario, 40 random topologies were simulated. As we aim to study high interference, the network operates in single channel to induce high interference in the network. In actual wireless networks which normally operate using multi-channel, high interferences only occur in each channel when the number of STAs increase in greater number than 36 used for our setup. The other parameters used in the simulation are shown in Table \ref{fig:parameter}. The traffic load is chosen such that the IFQ always have a packet to send. Some examples of the random topologies used in the simulation are shown in Fig. \ref{fig:toposeed} when OA are used and the number of STA is 9; the solid lines represent data links, the dashed lines represent nodes within receiving range, and the dotted lines represent nodes within carrier sensing range. As a node with directional antenna uses 4 interfaces, for correct comparison of aggregated throughout for a network using OA each node is fitted with 4 interfaces of OA. In practice only one interface will be active at any one time due to carrier sensing among interfaces.

\begin{table*}[ht] 
\caption{Parameter settings used in ns-2.33 simulations} 
\centering
\footnotesize
\begin{tabular}{r l} 
\hline 
Parameter & Setting\\ [0.5ex] 
\hline 
Access Scheme & Basic Access Scheme (DATA, ACK) \\ 
Rate & 11 Mbit/s (Data), 1 Mbit/s (Basic)\\ 
MAC & IEEE 802.11b\\
Offered Load & 55 packet/s/node \\
Traffic Packet Size & 1500 bytes \\
IFQ Length & 50 packets \\
Signal to Interference Ratio & 10 dB \\
Propagation & Two Ray Ground Reflection\\
Contention Window (CW) & 31 (Min), 1023 (Max) \\
ns-2's Default Transmit Power & 281.84 mW\\
Threshold & RX:3.65e-10 W, CS:1.79e-12 W\\
Traffic & UDP, Poisson process, 1818.181$\mu$s mean inter-arrival interval\\
Simulation Time & 120 s \\
Type of Antenna & OA, DA \\
Antenna Gain & OA:1, DA:2 \\
Number of DA/node & 4, 90$^{\circ}$ beamwidth each \\
Directional Antenna Angles & 0$^{\circ}$, 90$^{\circ}$, 180$^{\circ}$, 270$^{\circ}$ \\
Node Mobility & Static \\
Number of Simulations for Each Scenario & 40 \\
Number of STAs & 9, 18, 27, 36 \\ [1ex] 
\hline 
\end{tabular} 
\normalsize
\label{fig:parameter}
\end{table*} 

\subsection{Attacking Case}
We evaluate our improved $\textit{Attacking}$ $\textit{Case}$ metric against Liew's $\textit{Attacking}$ $\textit{Case}$ over a wireless network and compare the results of both. The value of $\textit{Attacking}$ $\textit{Case}$ indicates the potential for packet collisions and exponential backoffs in a wireless network; the higher the value of $\textit{Attacking}$ $\textit{Case}$ the smaller will be the aggregated throughput observed in the network.

\begin{figure}
\centering
\includegraphics[clip,width=0.8\columnwidth]{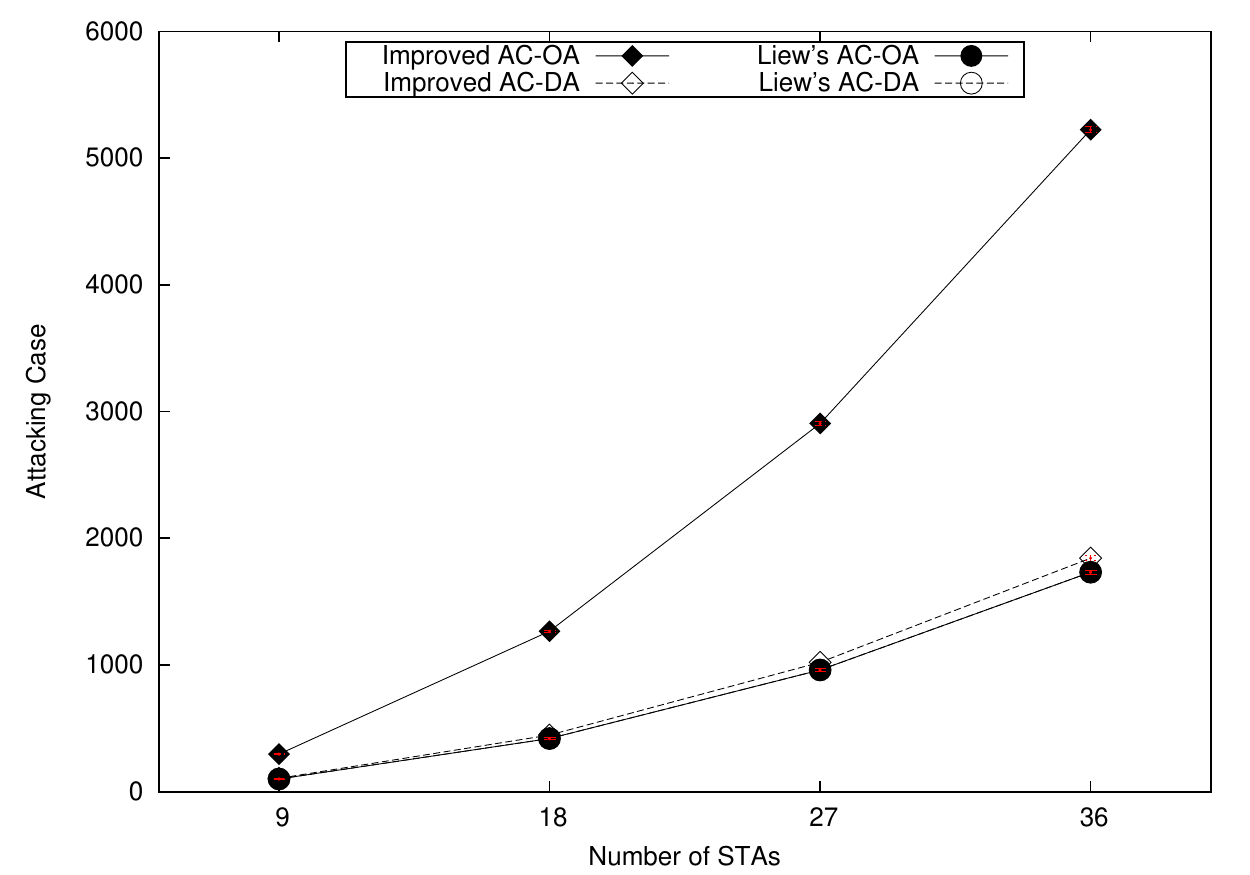}
\caption{The improved and Liew's $\textit{Attacking}$ $\textit{Case}$ metric for OA and DA when the number of STAs increase. Liew's AC-OA line overlaps with Liew's AC-DA line for all the number of STAs.} 
\label{fig:iGraph}
\end{figure}

\begin{figure}
\centering
\includegraphics[clip,width=0.8\columnwidth]{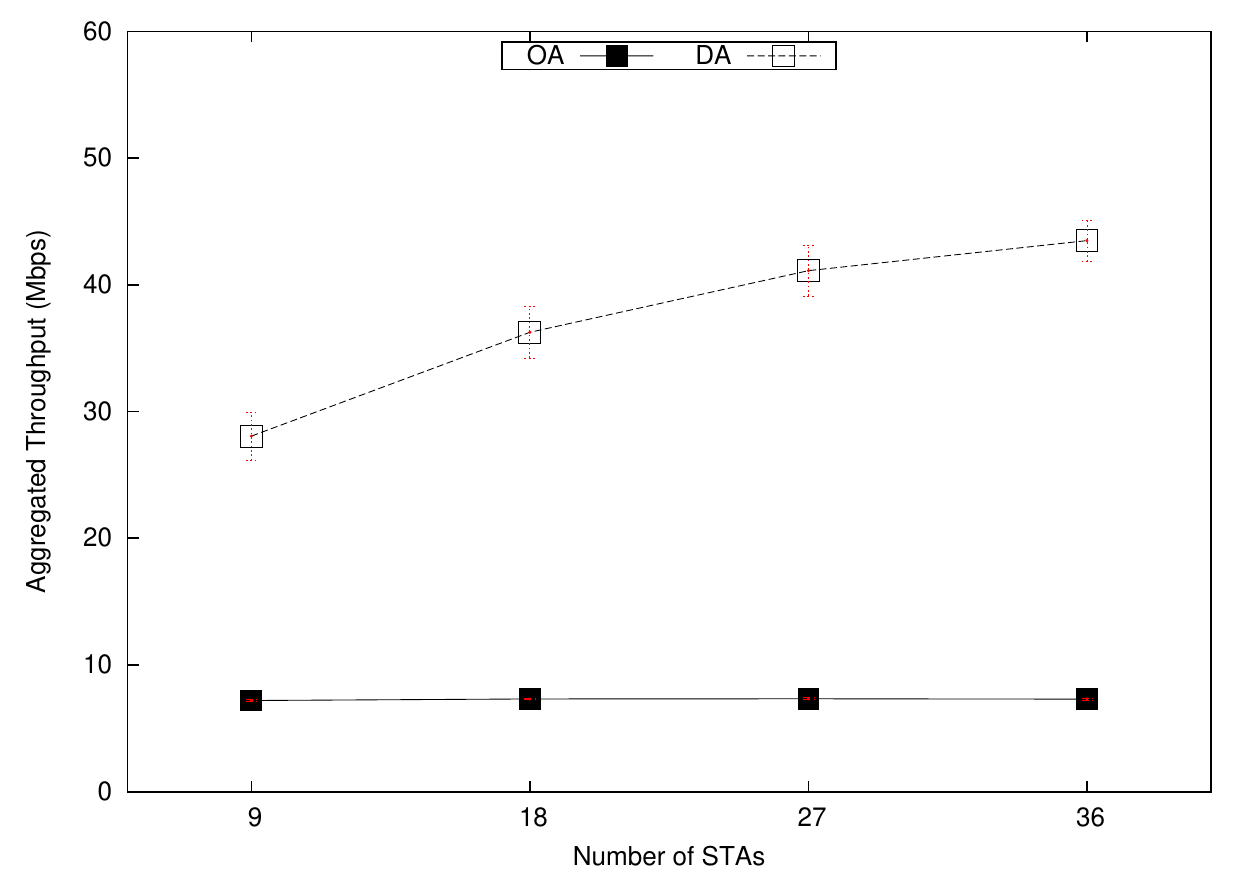}
\caption{The aggregated network throughput for OA and DA when the number of STAs increase.} 
\label{fig:sta_tput}
\end{figure}

Using the setup described in Section 5.2, the simulation results for $\textit{Attacking}$ $\textit{Case}$ for networks with nodes using OA and DA are presented in Fig. \ref{fig:iGraph}. The solid lines represent networks with nodes using OA and the dashed lines represent networks with nodes using DA. The x-axis captures the total number of STAs in the network. The number of STAs were increased by incrementing the STA/AP ratio (1, 2, 3, 4). On the y-axis, the $\textit{Attacking}$ $\textit{Case}$ in the network is calculated using our improved approach and Liew's approach. The simulation results for aggregated network throughput are also presented in Fig. \ref{fig:sta_tput} against the total number of STAs in the network. There are four curves in Fig. \ref{fig:iGraph} representing the improved and Liew's $\textit{Attacking}$ $\textit{Case}$ for OA and DA. In Fig. \ref{fig:sta_tput}, there are two curves for the aggregated throughput for network using nodes with OA and DA. 

\begin{figure}
\centering
\includegraphics[clip,scale=1.5]{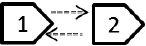}
\caption{I-graph using Liew's method for Network 2 with nodes using DA.} 
\label{fig:igraphivannet2}
\end{figure}

\subsubsection{Liew's Attacking Case and Directional Antenna}

Firstly, we show that the Liew's $\textit{Attacking}$ $\textit{Case}$ does not model adequately networks consisting of nodes using DA. In Fig. \ref{fig:iGraph} Liew's $\textit{Attacking}$ $\textit{Case}$ is presented by the lines with circle points. We can observe that the value of Liew's $\textit{Attacking}$ $\textit{Case}$ increases as the number of STAs increase due to the surge of interference. However the OA line overlaps with the DA line though the interference is reduced due to the capability of DA to reduce interference on unwanted directions. There are two reasons for this: a) weight of edges $w(e_{ij})$ - the Liew's $\textit{Attacking}$ $\textit{Case}$ metric is calculated using Eq. \ref{eq:19}. As the edge's weight is not considered in its calculation and only depends on the presence of an edge, the $\textit{Attacking}$ $\textit{Case}$ value for OA and DA is the same using Liew's approach. For Network 1 in Fig. \ref{fig:Graph SK Liew}, the $\textit{Attacking}$ $\textit{Case}$ calculated using Liew's method is 8 for both OA and DA; b) direction of transmission, $\theta$ - the $\textit{Attacking}$ $\textit{Case}$ calculated using Liew's method for Network 2 in Fig. \ref{fig:Graph SK Liew} is 8 for OA; this value considers the i-edges caused by ACK$_{2}$-DATA$_{1}$ and ACK$_{1}$-DATA$_{2}$ pairs of transmissions. For DA though the i-edges due to ACK$_{2}$-DATA$_{1}$ and ACK$_{1}$-DATA$_{2}$ are no longer present because the DA is able to point its beam to its intended direction and reduce interference on unwanted direction, but since $\theta$ was not considered by Liew for the construction of the power constraints the same i-graph would result for DA and OA. The resultant i-graph for Network 2 using Liew's method is shown in Fig. \ref{fig:igraphivannet2}. Thus the $\textit{Attacking}$ $\textit{Case}$ value for DA will be the same as OA. In conclusion, the $\textit{Attacking}$ $\textit{Case}$ metric calculated by Liew gives the same value for OA and DA irrespective of the number of STAs, as shown in Fig. \ref{fig:iGraph}. However when the aggregated network throughput of OA and DA is evaluated in Fig. \ref{fig:sta_tput} there are big differences between them. DA's throughput outperforms OA by at least 290\% for the case of 9 STAs, calculated according to Eq. \ref{eq:20}. This suggests Liew's $\textit{Attacking}$ $\textit{Case}$ is not adequate to quantize the severity of interference for networks with nodes using DA.
\small
\begin{equation} \label{eq:20}
Gain = \left(Tput_{DA}-Tput_{OA}\right) \times 100/Tput_{OA}
\end{equation}
\normalsize

\subsubsection{Improved Attacking Case supporting Directional Antenna}

Secondly we show that our improved $\textit{Attacking}$ $\textit{Case}$ supports nodes using DA and it is also compatible with nodes using OA. In Fig. \ref{fig:iGraph} the improved $\textit{Attacking}$ $\textit{Case}$ is presented by the lines with diamond shaped points. It can be observed that the value of $\textit{Attacking}$ $\textit{Case}$ increases as the number of STAs increase for both OA and DA. The OA increases with higher slopes than DA. It can also be seen that the improved $\textit{Attacking}$ $\textit{Case}$ no longer causes overlapping lines between OA and DA. This is because the weight of edges $w(e_{ij})$ and direction of transmission $\theta$ considered in our method are important parameters to characterize the interference caused by nodes using DA. When the number of STAs is 36, the improved $\textit{Attacking}$ $\textit{Case}$ for OA is approximately 5220 and when the DA setup is used the value decreases to 1840, showing the potential high gain foreseeable in throughput. This is confirmed by the throughput lines in Fig. \ref{fig:sta_tput} where DA performed close to 500\% better than the OA for the case of 36 STAs. 

In Fig. \ref{fig:sta_tput}, as the number of STAs increase the aggregated throughput for DA increases but the rate of increase reduces. This is because the network with nodes using DA is getting saturated. Adding more STAs though increase the amount of offered load to the network unfortunately the network unable to transport more packets due to high exponential backoffs and collisions persist in the network. For OA, due to the nature of the antenna transmitting at all direction, the network gets saturated at much lower STAs than DA as shown in Fig. \ref{fig:sta_tput}. Due to this reason the aggregated throughput is constant for OA even though the attacking case in Fig. \ref{fig:iGraph} increases.

\subsubsection{Using Improved Attacking Case in Networks with Various Transmission Power}
Thirdly, we show that the improved $\textit{Attacking}$ $\textit{Case}$ metric is useful to quantize the severity of interference in networks where various transmission powers are used. Let us define the default transmission power in ns-2 as DP-NChan \cite{ns-2}. In order to evaluate different levels of interference and its effect on $\textit{Attacking}$ $\textit{Case}$, apart from using DP-NChan, the network is also simulated using a minimum transmit power (MP) approach. In this approach the transmission power is enough for a transmitter node to get its transmitted packets decoded by its receiving node. We studied the minimum transmit power approach by using the following 3 setups:

\begin{itemize}
  \item the minimum power per network (MP-PNetw) -- in this setup the interfaces in nodes are allowed to reduce its transmission power, but all the interfaces in the network must use the same transmission power. OA and DA use it.
  \item the minimum power per node (MP-PNode) -- in this setup, as above, the interfaces are allowed to reduce its transmission power. Each node is allowed to have its own transmission power but all the interfaces of a node must use the same power. OA and DA use it.
  \item the minimum power per interface (MP-PInte) -- in this setup each interface is allowed to reduce and use its own transmission power. Only DA uses this. 
\end{itemize}

\begin{figure}
\centering
\includegraphics[clip,width=\columnwidth]{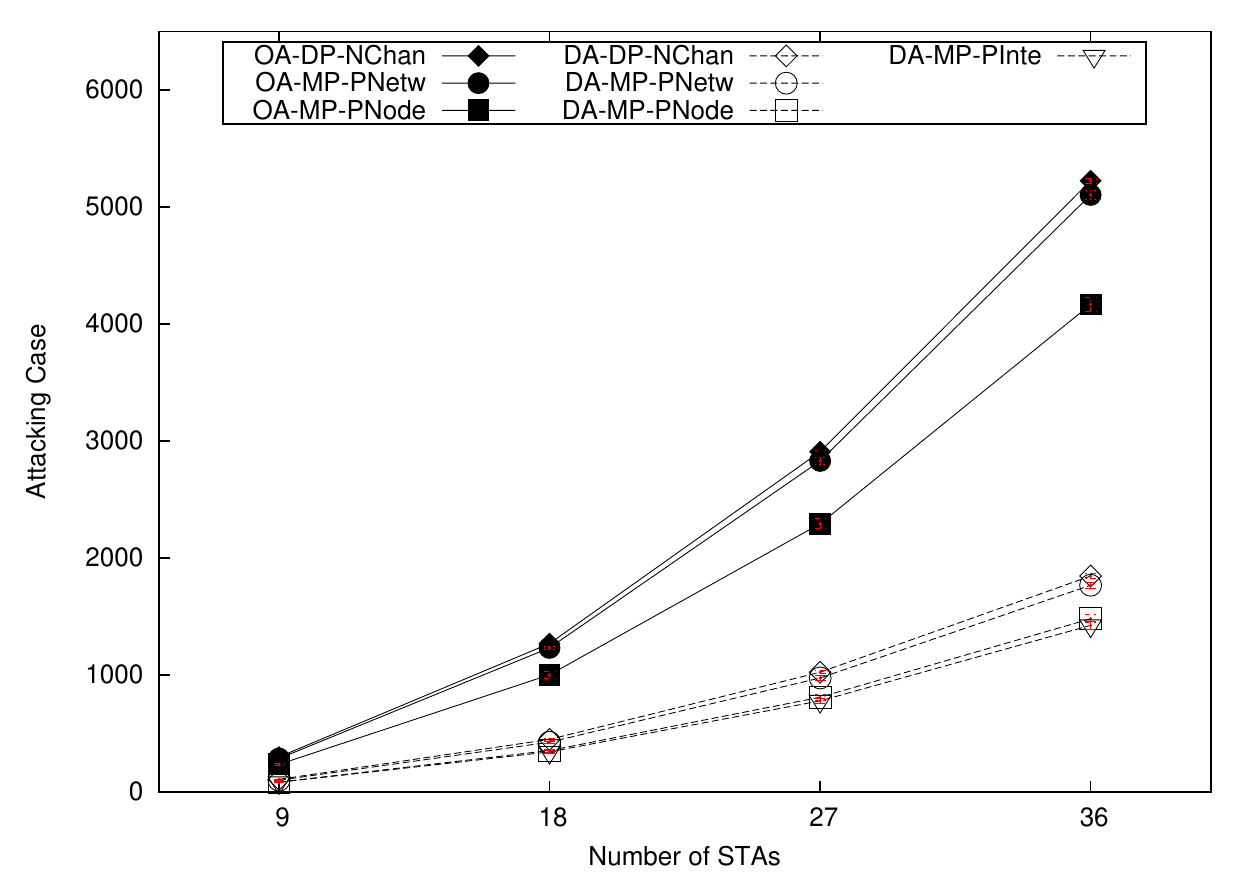}
\caption{The improved $\textit{Attacking}$ $\textit{Case}$ metric for OA and DA for various transmission power strategies when the number of STAs increase.} 
\label{fig:iiap}
\end{figure}

\begin{figure}
\centering
\includegraphics[clip,width=\columnwidth]{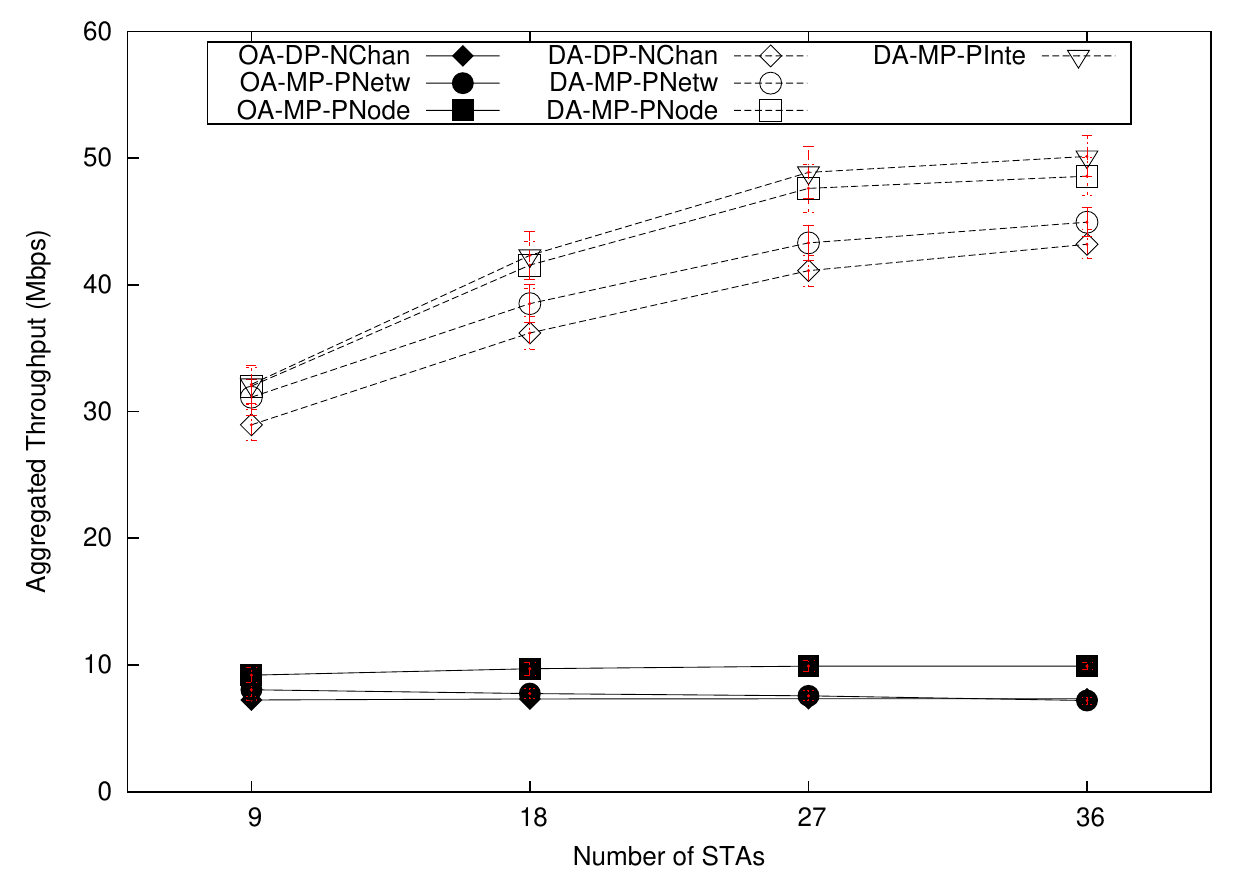}
\caption{The aggregated network throughput for OA and DA for various transmission power strategies when the number of STAs increase.} 
\label{fig:tputap}
\end{figure}

\begin{figure}
\centering
\includegraphics[clip,width=\columnwidth]{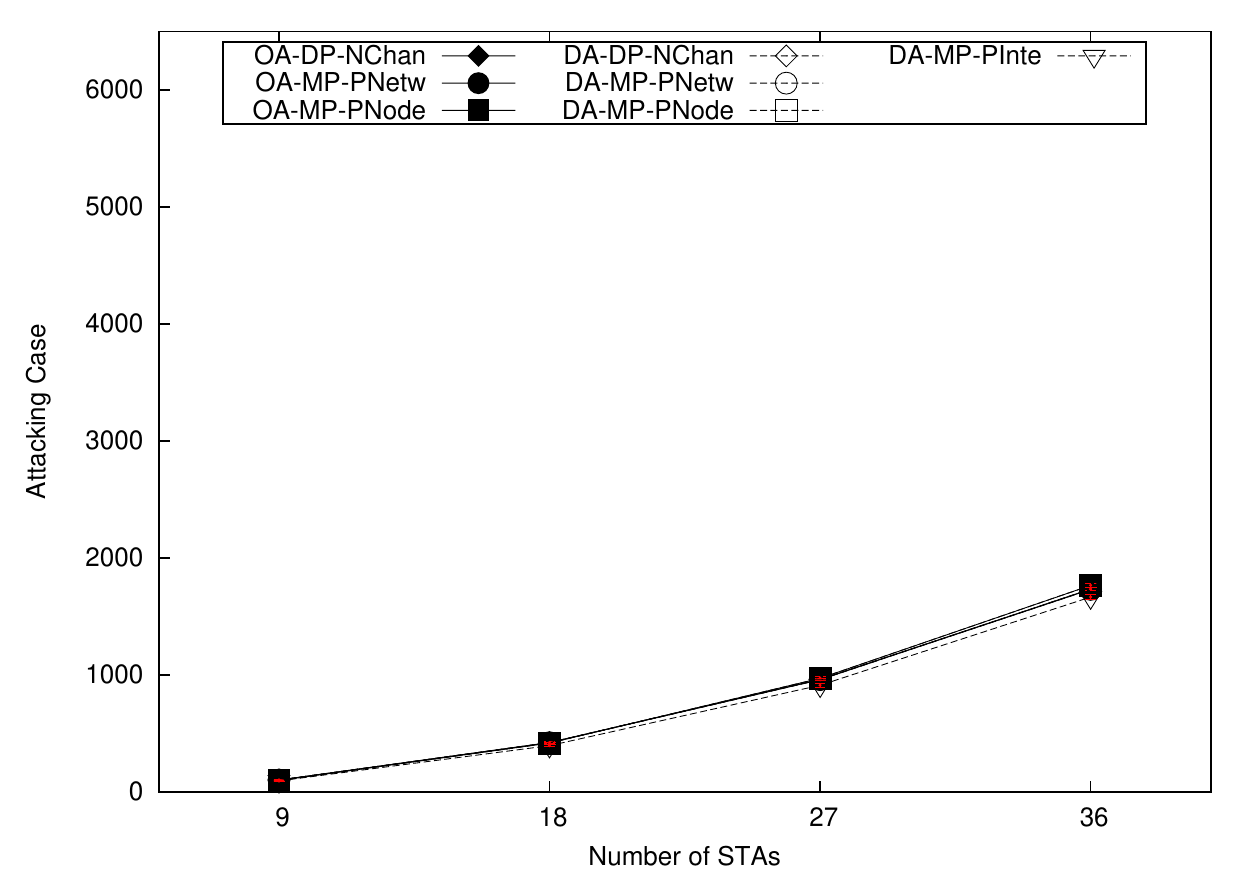}
\caption{Liew's $\textit{Attacking}$ $\textit{Case}$ metric for OA and DA for various transmission power strategies when the number of STAs increase.} 
\label{fig:liewac}
\end{figure}

The rest of the parameters used for the simulations are shown in Table \ref{fig:parameter}. The improved $\textit{Attacking}$ $\textit{Case}$ and Liew's $\textit{Attacking}$ $\textit{Case}$ were calculated using Eq. \ref{eq:17} and Eq. \ref{eq:19} respectively for all these networks. The simulation results are shown in Fig. \ref{fig:iiap}, Fig. \ref{fig:tputap} and Fig. \ref{fig:liewac}.

Fig. \ref{fig:iiap} shows the graph for improved $\textit{Attacking}$ $\textit{Case}$ versus the number of STAs in the network. Solid lines represent networks with nodes using OA while dashed lines represent networks with nodes using DA. As the number of STAs increases, the amplitude of $\textit{Attacking}$ $\textit{Case}$ increases for all the setups. When minimum transmission power approach is used, the $\textit{Attacking}$ $\textit{Case}$ for the 3 setups is reduced compared with the default transmission power setup for both OA and DA. For example, for the network with 36 STAs the $\textit{Attacking}$ $\textit{Case}$ is reduced by 22\% for network using DA with minimum transmit power per interface setup compared with DA using default transmit power setup. This is because the transmission power reduction assists to reduce the amount of interference in the network. When comparing the 3 minimum transmit power setups we can observe, as expected, that the minimum transmit power per interface is the most attractive setup followed by minimum transmit power per node, and minimum transmit power per network. This is well captured by the improved $\textit{Attacking}$ $\textit{Case}$ metric. 

Fig. \ref{fig:tputap} represents the aggregated throughput versus the number of STAs in the network. Solid lines represent networks with nodes using OA while dashed lines represent networks with nodes using DA. As the number of STAs increases the throughput is constant for OA but DA has higher throughput although the slope of the throughput line decreases for all the setups. The throughput is higher when minimum transmission power approach is used for both type of antennas. For example, for the network with 36 STAs the throughput observed for the network using DA with minimum transmit power per interface setup is 16\% higher than the throughput obtained on the equivalent network with DA using default transmit power. This is because the transmission power reduction reduces interference in the network as reflected by the improved $\textit{Attacking}$ $\textit{Case}$ metric in Fig. \ref{fig:iiap} and this allows more packets to be transmitted per second. For OA, minimum transmit power per network has no significant throughput gain than default transmit power. This is because the transmission power reduction approach is unable to reduce sufficient interference as shown in Fig. \ref{fig:iiap}. Hence the throughput did not increase greatly. Nevertheless, for the network with 36 STAs the throughput observed for the network using OA with minimum transmit power per node setup is 35\% higher than the throughput obtained on the equivalent network with OA using default transmit power.

Fig. \ref{fig:liewac} shows the graph for Liew's $\textit{Attacking}$ $\textit{Case}$ versus the number of STAs in the network. Solid lines represent networks with nodes using OA while dashed lines represent networks with nodes using DA. We can observe that the value of Liew's $\textit{Attacking}$ $\textit{Case}$ increases with the increase of the number of STAs due to the higher accumulation of interference in the network. However the lines of various transmission power setups are similar with one another suggesting all these setups have the same severity of interference in the network and potentially lead to similar aggregated throughput. But the result in Fig. \ref{fig:tputap} shows the various transmission power setups majorly have different aggregated throughput. 

Liew's $\textit{Attacking}$ $\textit{Case}$ in Eq. \ref{eq:19} consists of components in Eq. \ref{eq:19i}, Eq. \ref{eq:19tc} and Eq. \ref{eq:19rc}. Table \ref{fig:liewedge} shows the values for these components for the example of network with 36 STAs. As the sum of i-edges, and rc-edges that are not part of tc-edges and i-edges increases, the sum of tc-edges that are not part of i-edges reduces at similar rate. This causes similar $\textit{Attacking}$ $\textit{Case}$ values irrespectively of the transmission power reduction approach used for each type of antenna. Using Liew's method, only edges that are not in Component 1 will be considered for Component 2 and Component 3. Since Component 2 taken most of the remaining edges, Component 3 left with few edges as shown in Table \ref{fig:liewedge}.
\small
\begin{equation} \label{eq:19i} 
\begin{split} 
\smash[b]{\sum_{\substack{i,j \in \mathit{V}\\i\neq j}}}
\;\bigl[ \mathds{1}_{[e_{i,j} \in E_{I}]} \bigr]
\end{split} 
\end{equation} 
\begin{equation} \label{eq:19tc} 
\begin{split} 
\smash[b]{\sum_{\substack{i,j \in \mathit{V}\\i\neq j}}}
\;\bigl[ \mathds{1}_{[e_{i,j} \in E_{TC}\ \wedge\ e_{i,j} \notin E_{I}]} \bigr]
\end{split} 
\end{equation} 
\begin{equation} \label{eq:19rc} 
\begin{split} 
\smash[b]{\sum_{\substack{i,j \in \mathit{V}\\i\neq j}}}
\;\bigl[ \mathds{1}_{[e_{i,j} \in E_{RC}\ \wedge\ e_{i,j} \notin E_{TC}\ \wedge\ e_{i,j} \notin E_{I}]} \bigr]
\end{split} 
\end{equation} 
\normalsize
\begin{table}[ht] 
\caption{The components of Eq. \ref{eq:19} and the resultant $\textit{Attacking}$ $\textit{Case}$ using Liew's method when the number of STAs is 36}
\centering
\footnotesize
\begin{tabular}{cccccc} 
\hline 
\multirow{2}{*}{Method}	&	\multirow{2}{*}{Setup} & Component 1      & Component 2      & Component 3      &	\multirow{2}{*}{AC$_{Liew}$}\\ 
                 	      &	    	                 & (Eq. \ref{eq:19i}) & (Eq. \ref{eq:19tc}) & (Eq. \ref{eq:19rc}) &	\\ [0.5ex] 
\hline 
\\\multirow{3}{*}{OA}	&	DP-Nchan	&	472.1	&	787.9	&	0.0	&	1732.1	\\
	&	MP-PNetw	&	472.1	&	763.8	&	21.8	&	1729.8	\\
	&	MP-PNode	&	549.1	&	614.7	&	51.3	&	1764.1	\\ [1ex]		
\hline 
\\\multirow{4}{*}{DA}	&	DP-Nchan	&	472.1	&	787.9	&	0.0	&	1732.1	\\		
	&	MP-PNetw	&	472.1	&	763.8	&	21.8	&	1729.8	\\		
	&	MP-PNode	&	549.1	&	614.7	&	51.3	&	1764.1	\\
	&	MP-Pinte	&	518.1	&	564.0	&	64.6	&	1664.7	\\[1ex]		
\hline 
\end{tabular} 
\normalsize
\label{fig:liewedge} 
\end{table} 

The improved $\textit{Attacking}$ $\textit{Case}$ in Eq. \ref{eq:17} consists of components in Eq. \ref{eq:17i}, Eq. \ref{eq:17tc} and Eq. \ref{eq:17rc}. Table \ref{fig:improvededge} shows the values of these components for the same network. It shows dissimilar values of $\textit{Attacking}$ $\textit{Case}$ compare with Table \ref{fig:liewedge}. The improved $\textit{Attacking}$ $\textit{Case}$ metric is able to represent the changes in the aggregated throughput in Fig. \ref{fig:tputap} more accurately. This shows the usage of weight of edges $w(e_{ij})$ is important to model the severity of interference in networks where various transmission powers are used.
\small
\begin{equation} \label{eq:17i} 
\begin{split} 
\smash[b]{\sum_{\substack{i,j \in \mathit{V}\\i\neq j}}}
\;\bigl[ w_{I}(e_{i,j}) \times \mathds{1}_{[e_{i,j} \in E_{I}]} \bigr]
\end{split} 
\end{equation} 
\begin{equation} \label{eq:17tc} 
\begin{split} 
\smash[b]{\sum_{\substack{i,j \in \mathit{V}\\i\neq j}}}
\;\bigl[ w_{TC}(e_{i,j}) \times \mathds{1}_{[e_{i,j} \in E_{TC}\ \wedge\ e_{i,j} \notin E_{I}]} \bigr]
\end{split} 
\end{equation} 
\begin{equation} \label{eq:17rc} 
\begin{split} 
\smash[b]{\sum_{\substack{i,j \in \mathit{V}\\i\neq j}}}
\;\bigl[ w_{RC}(e_{i,j}) \times \mathds{1}_{[e_{i,j} \in E_{RC}\ \wedge\ e_{i,j} \notin E_{I}]} \bigr]
\end{split} 
\end{equation}
\normalsize

\begin{table}[ht] 
\caption{The components of Eq. \ref{eq:17} and the resultant $\textit{Attacking}$ $\textit{Case}$ using Improved method when the number of STAs is 36} 
\centering 
\footnotesize
\begin{tabular}{cccccc}
\hline 
\multirow{2}{*}{Method}	&	\multirow{2}{*}{Setup} & Component 1      & Component 2      & Component 3      &	\multirow{2}{*}{AC$_{Imp}$}\\ 
                 	      &	    	                 & (Eq. \ref{eq:17i}) & (Eq. \ref{eq:17tc}) & (Eq. \ref{eq:17rc}) &	\\ [0.5ex] 
\hline 
\\\multirow{3}{*}{OA}	  &	DP-Nchan	&	1040.7	&	1573.1	&	1569.8	&	5224.1	\\
	&	MP-PNetw	&	1040.7	&	1484.1	&	1536.7	&	5102.2	\\
	&	MP-PNode	&	1052.8	&	993.3	&	1066.7	&	4165.6	\\[1ex]	
\hline  
\\\multirow{4}{*}{DA}	&	DP-Nchan	&	337.4	&	537.2	&	631.3	&	1843.3	\\
	&	MP-PNetw	&	337.4	&	489.4	&	601.0	&	1765.2	\\
	&	MP-PNode	&	332.7	&	362.7	&	450.8	&	1478.8	\\
	&	MP-Pinte	&	329.8	&	338.4	&	424.8	&	1422.6	\\[1ex]	
\hline 
\end{tabular} 
\normalsize
\label{fig:improvededge} 
\end{table} 

Comparing the 3 minimum transmit power setups, the power control per interface has the least interference in the network and, as a consequence, it leads to the highest aggregated network throughput. Then it is followed by power control per node, and power control per network. The default transmission approach is the least attractive setup. The additional degree of controlling power by interface in DA makes it more attractive than OA. In conclusion reducing $\textit{Attacking}$ $\textit{Case}$ can result in a potentially increase of throughput. The reduction of $\textit{Attacking}$ $\textit{Case}$ can be achieved by using strategies such as DA, transmission power reduction, or DA with transmission power reduction. 

We have shown that Liew's $\textit{Attacking}$ $\textit{Case}$ metric is not adequate for networks with nodes using DA; hence the need for a new $\textit{Attacking}$ $\textit{Case}$ metric. We have also shown that our improved $\textit{Attacking}$ $\textit{Case}$ supports nodes using DA and it is compatible with nodes using OA; the improved $\textit{Attacking}$ $\textit{Case}$ metric is able to distinguish the severity of interference by network using nodes with DA and OA. Lastly, we have shown that our improved $\textit{Attacking}$ $\textit{Case}$ can be used to quantize the interference in networks that use various transmission power schemes.

\section{Conclusions}
Interference is a fundamental issue in wireless networks and it affects the aggregated throughput of a network. In this paper we have characterized the power interference in IEEE 802.11 CSMA/CA based networks using DA. An improved $\textit{Attacking}$ $\textit{Case}$ metric that quantizes the severity of interference has been proposed using the Link-Interference Graph, Transmitter-side Protocol Collision Prevention Graph, and Receiver-side Protocol Collision Prevention Graph. This metric differs from Liew's $\textit{Attacking}$ $\textit{Case}$ metric proposed in \cite{ivan} as the original metric only addresses networks using OAs. Our improved $\textit{Attacking}$ $\textit{Case}$ metric is meant for networks using DA but it can also be used in networks using OA. It was also found that interference is tied with $\textit{Attacking}$ $\textit{Case}$, thus reducing $\textit{Attacking}$ $\textit{Case}$ can result in an increase of throughput. The reduction of $\textit{Attacking}$ $\textit{Case}$ can be achieved by the usage of strategies such as DA, transmission power reduction, or DA with transmission power reduction. The relationship between $\textit{Attacking}$ $\textit{Case}$ and the throughput of a network is worth to be studied; if there is a statistically strong relationship between these two, a model could be built which is useful to predict the throughput of a network once its $\textit{Attacking}$ $\textit{Case}$ is calculated. The prediction model would be of assistance in the planning process of a network. This activity remains as our future work. 

It would be advantageous to use $\textit{Attacking}$ $\textit{Case}$ to predict the throughput as the $\textit{Attacking}$ $\textit{Case}$ metric could be calculated using simple procedure with the knowledge of node positions, transmission power, signal to interference ratio and radio propagation rather than using a discrete event network simulator. Network simulators demand simulator specific codes to be developed, multiple simulations to be executed, wait for the simulations to be completed, and output logs to be analysed; only then one would have the knowledge on the expected throughput. 

\section{Acknowledgments}
The authors would like to thank the Fundação para a Ciência e a Tecnologia (FCT) of Ministério da Ciência, Tecnologia e Ensino Superior (MCTES), Portugal for supporting this work through grant SFRH/BD/43744/2008 and PTDC/EEA-TEL/120176/2010.

\bibliography{mybiblo}

\vspace{10mm}

\parpic{\includegraphics[width=1in,height=1.25in,clip,keepaspectratio]{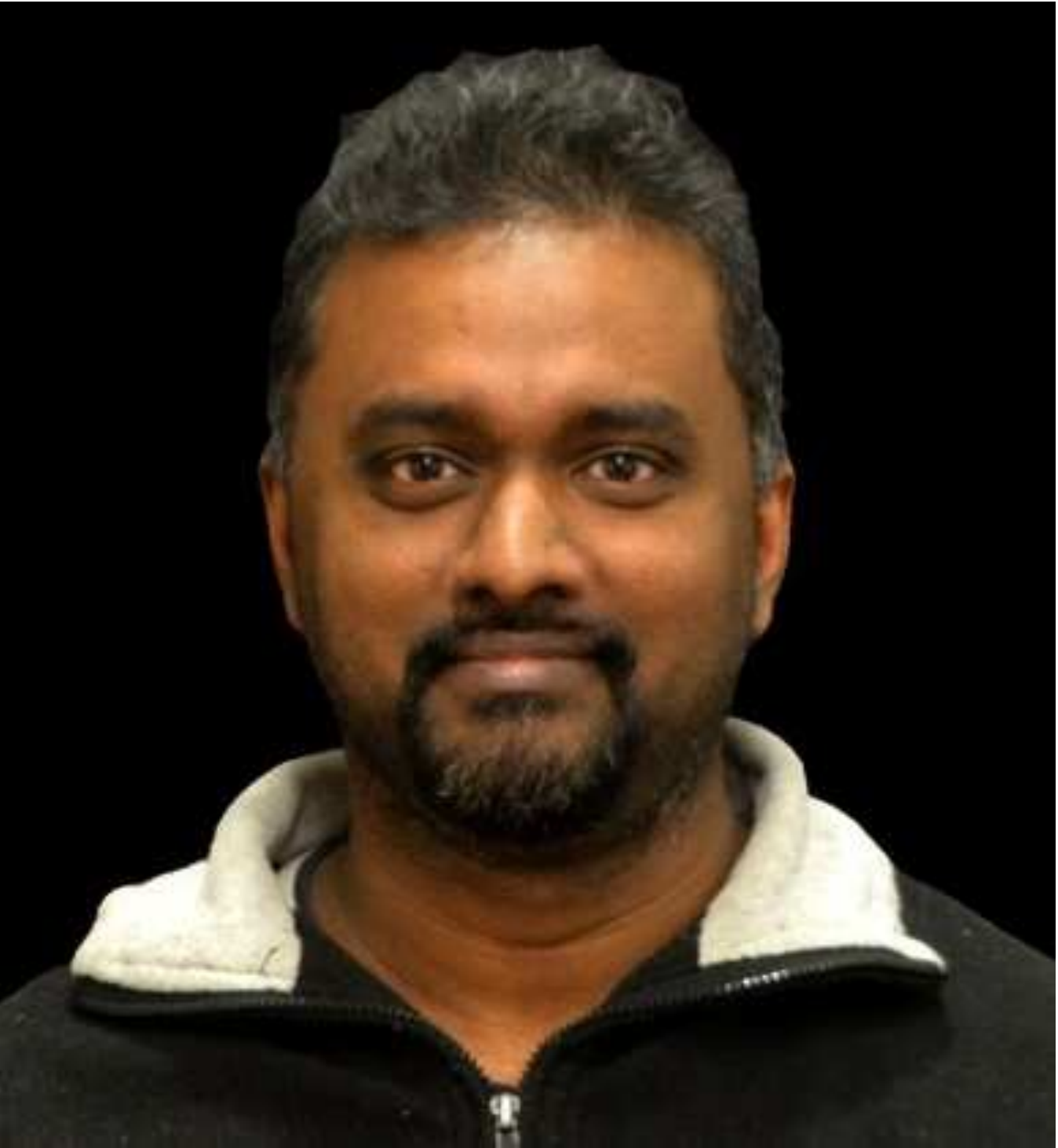}}
\noindent {\bf Saravanan Kandasamy} received the B.Eng (2000) in Electronics majoring in Computer from Multimedia University, Malaysia and M.Sc (2004) in Communications and Network Engineering from University Putra Malaysia. He is currently a researcher in the Centre for Telecommunications and Multimedia of INESC TEC (wwww.inesctec.pt) and pursuing Ph.D in the MAP Doctoral Programme (www.tele.map.edu.pt). His research interest include directional antenna, radio resource management, transmission power control and statistical modeling for IEEE 802.11 based wireless networks.\\ \\

\parpic{\includegraphics[width=1in,height=1.25in,clip,keepaspectratio]{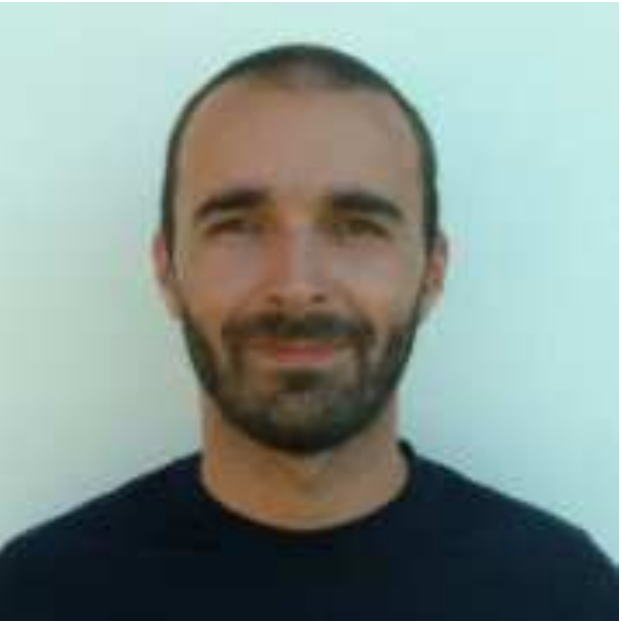}}
\noindent {\bf Ricardo Morla} is an Assistant Professor with the Electrical and Computer Engineering Department, and a principal investigator with INESC Porto, at the Faculty of Engineering of the University of Porto. His research interests are in the field of modeling and management of IT systems with an emphasis on probabilistic and machine learning approaches applied to networks and ambient intelligence. Ricardo graduated from U.Porto in Electrical and Computer Engineering and holds a PhD in Computing from Lancaster University. He was a lecturer and post-doc at UC Irvine in 2007, and a visiting faculty at Carnegie Mellon University in 2010 under the CMU-Portugal program. \\ \\

\parpic{\includegraphics[width=1in,height=1.25in,clip,keepaspectratio]{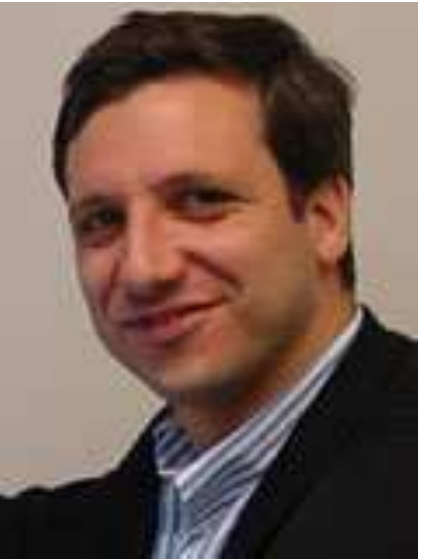}}
\noindent {\bf Manuel Ricardo} received a Licenciatura (1988), M.Sc (1992), and PhD (2000) degrees in Electrical and Computer Engineering from Porto University. Currently, he is an associate professor at the Faculty of Engineering of University of Porto, where he gives courses in mobile communications and computer networks. He also coordinates Centre for Telecommunications and Multimedia of INESC TEC (wwww.inesctec.pt).

\end{document}